\documentclass[twocolumn]{aastex63}

\newcommand{\borus}{\texttt{Borus}}
\newcommand{\mytorus}{\texttt{MYTorus}}
\newcommand{\xspec}{\textsc{xspec}}

\newcommand{\nustar}{\textit{NuSTAR}}

\newcommand{\mbh}{$M_\mathrm{BH}$}

\usepackage{booktabs}
\usepackage{multirow}
\usepackage{array}
\usepackage{hyperref}
\newcolumntype{H}{>{\setbox0=\hbox\bgroup}c<{\egroup}@{}}
\usepackage{efbox,graphicx}
\usepackage{stackengine,xcolor}
\efboxsetup{linecolor=black,linewidth=1pt}
\usepackage[caption=false]{subfig}
\definecolor{nsgreen}{rgb}{0.1,0.5,0.1}

\shorttitle{FRAMEx III: High-sensitivity VLBA observation of radio quiet AGNs}
\shortauthors{Shuvo et al.}
\graphicspath{{./}{figures/}}

\begin{document}

\title{Fundamental Reference AGN Monitoring Experiment (FRAMEx) III:\\[0.05cm] Radio Emission in the Immediate Vicinity of Radio Quiet AGNs}

\correspondingauthor{Onic Islam Shuvo}
\email{oshuvo@gmu.edu}

\author[0000-0003-4727-2209]{Onic I. Shuvo}
\affiliation{U.S. Naval Observatory, 3450 Massachusetts Ave NW, Washington, DC 20392-5420, USA}
\affiliation{Department of Physics and Astronomy, George Mason University, MS3F3, 4400 University Drive, Fairfax, VA 22030, USA}

\author[0000-0002-4146-1618]{Megan C. Johnson}
\affiliation{U.S. Naval Observatory, 3450 Massachusetts Ave NW, Washington, DC 20392-5420, USA}

\author[0000-0002-4902-8077]{Nathan J. Secrest}
\affiliation{U.S. Naval Observatory, 3450 Massachusetts Ave NW, Washington, DC 20392-5420, USA}

\author[0000-0002-8818-9009]{Mario Gliozzi}
\affiliation{Department of Physics and Astronomy, George Mason University, MS3F3, 4400 University Drive, Fairfax, VA 22030, USA}

\author[0000-0002-3365-8875]{Travis C. Fischer}
\affiliation{AURA for ESA, Space Telescope Science Institute, Baltimore, MD, USA, 3700 San Martin Drive, Baltimore, MD 21218, USA}

\author[0000-0002-8736-2463]{Phillip J. Cigan}
\affiliation{Department of Physics and Astronomy, George Mason University, MS3F3, 4400 University Drive, Fairfax, VA 22030, USA}
\affiliation{U.S. Naval Observatory, 3450 Massachusetts Ave NW, Washington, DC 20392-5420, USA}

\author[0000-0002-0819-3033]{Luis C. Fernandez}
\affiliation{U.S. Naval Observatory, 3450 Massachusetts Ave NW, Washington, DC 20392-5420, USA}
\affiliation{Department of Physics and Astronomy, George Mason University, MS3F3, 4400 University Drive, Fairfax, VA 22030, USA}

\author[0000-0002-5604-5254]{Bryan N. Dorland}
\affiliation{U.S. Naval Observatory, 3450 Massachusetts Ave NW, Washington, DC 20392-5420, USA}



\begin{abstract}

We present follow-up results from the first Fundamental Reference AGN Monitoring Experiment (FRAMEx) X-ray/radio snapshot program of a volume-complete sample of local hard X-ray-selected active galactic nuclei (AGNs). 
Here, we added 9 new sources to our previous volume-complete snapshot campaign, two of which are detected in the 6 cm Very Long Baseline Array (VLBA) observations. We also obtained deeper VLBA observations for a sample of 9 AGNs not detected by our previous snapshot campaign. We recovered 3 sources with approximately twice the observing sensitivity. In contrast with lower angular resolution Very Large Array (VLA) studies, the majority of our sources continue to be undetected with the VLBA. The sub-parsec radio (6~cm) and X-ray ($2-10$~keV) emission shows no significant correlation, with~$L_\mathrm{R}/L_\mathrm{X}$ ranging from $10^{-8}$ to $10^{-4}$, and the majority of our sample lies well below the fiducial $10^{-5}$ relationship for coronal synchrotron emission. Additionally, our sources are not aligned with any of the proposed ``fundamental'' planes of black hole activity, which purport to unify black hole accretion in the $M_\mathrm{BH}-L_\mathrm{X}-L_\mathrm{R}$ parameter space. The new detections in our deeper observations suggest that the radio emission may be produced by the synchrotron radiation of particles accelerated in low luminosity outflows. Non-detections may be a result of synchrotron self-absorption at 6~cm in the radio core, similar to what has been observed in X-ray binaries (XRBs) transitioning from the radiatively inefficient state to a radiatively efficient state. 


\end{abstract}


\section{Introduction} \label{sec:intro}

The Fundamental Reference AGN Monitoring Experiment, or FRAMEx, is a research collaboration between the U.S. Naval Observatory (USNO) and other institutions that aims to monitor and characterize the physical properties of Active Galactic Nuclei (AGNs) powered by accretion onto supermassive black holes (SMBHs) at the centers of galaxies. To understand the physical nature and features of SMBHs, their surrounding media, and mutual interactions with the host galaxies that influence their luminosity and variability, FRAMEx used ground and space-based telescopes to observe AGNs in the X-ray and radio wavelengths at multiple time epochs~\citep{Dorland_2020jsrs.conf..165D}. 

Periods of AGN activity have a profound impact on the nuclear environment of galaxies, heating, ionizing, and blowing out gas and dust, regulating star-formation and gas accumulation. This feedback process in turn regulates the accretion rates of AGNs, leading to a positive correlation between the masses of SMBHs and their host galaxy stellar bulges over cosmic time \citep[for a review, see][]{Kormendy_2013ARA&A..51..511K}. Despite the substantial progress that has been made in recent years, the accretion mechanism of AGNs is still an active area of research, in part due to the wide range of morphologies and spectral energy distributions that AGNs exhibit, from compact, thermally-dominant quasars to large elliptical galaxies displaying jets and radio lobes extending to Mpc scales. Correlations between emission mechanisms at various wavelengths and properties of the black hole itself, such as the mass, have led to the exciting prospect that black holes exhibit self-similar accretion properties for all masses, from stellar-mass black holes in X-ray binaries (XRBs), to billion-plus \(\textup{M}_\odot\) SMBHs at the centers of the largest galaxies. A notable attempt to unify black hole accretion is the ``Fundamental Plane of Black Hole Activity'' \citep[e.g.,][hereafter the FP]{Merloni_2003MNRAS.345.1057M,Gultekin_2009ApJ...706..404G}, which purports to place all black holes in a single accretion parameter space, with one axis being X-ray emission, another being radio, and the third the black hole mass. The apparent FP relation from the previous lower angular resolution studies needs to be investigated further to understand the radiative processes and the physical environment very close to the black hole using observations at finer physical scales. 

To achieve this goal, in \citet[][hereafter, Paper~I]{Fischer_2021ApJ...906...88F}, we obtained simultaneous Swift X-ray Telescope (XRT) and Very Long Baseline Array (VLBA) radio observations for a snapshot (1-hour on-source) survey of 25 nearby AGNs ($<40$~Mpc) making up a volume-complete ($L_\mathrm{14-195~keV}>10^{42}$~erg~s$^{-1}$) sample at our declination range from $-30\arcdeg < \delta < +60\arcdeg$. One of the surprising results from Paper~I was that, despite being at the same radio frequency (C-band: 6~cm) and X-ray energies (2--10~keV) that the 
fundamental plane of black hole activity was defined at, we found that the FP breaks down at the angular resolution of the VLBA ($\sim3$~mas), calling into question its validity. Archival VLA data show that these objects do align with the FP at lower angular resolution ($\sim500$~mas, or $\gtrsim30-100$~pc), suggesting that whatever is responsible for the apparent correlation between SMBH mass and X-ray/radio emission paradoxically occurs on larger physical scales, at least beyond the largest angular scale of the VLBA observations ($\sim50$~mas, or $\gtrsim3-10$~pc for the typical distances of the sample), far larger than the scales of the AGN corona where X-ray emission is generally considered to arise. On the other hand, for radio loud (RL) AGNs, X-ray emission could be a superposition of different components such as synchrotron radiation from a jet, synchrotron self-Compton, emission from the accretion flow and inverse Compton scattering of lower energy photons off a corona \citep{Plotkin_2012MNRAS.419..267P}, but for our sample AGNs, which are almost entirely radio-quiet (RQ) (the exception is NGC~1052), the most likely explanation for the X-ray emission is the accretion disk corona and so far no current or planned future missions can resolve the X-ray emitting corona region in AGNs \citep[although see][]{2021ExA....51.1081U}. Despite the proximity of AGNs in the volume-complete sample and their selection at hard X-rays \citep{Oh_2018ApJS..235....4O}, only 9 out of the 25 AGNs were detected in the VLBA observations with a sensitivity level of $\sim20$~$\mu$Jy bm$^{-1}$ from Paper~I, raising the prospect of true ``radio-silent'' AGNs. 

In this paper, we present results from a follow-up VLBA observing campaign at C-band to observe 9 out of 16 initially non-detected AGNs with a much deeper sensitivity of $\sim10$~$\mu$Jy bm$^{-1}$ (4 hours on-source per target), to probe whether or not we were sensitivity limited in our initial snapshot campaign. Moreover, we added 9 new snapshot observations similar to our Paper I campaign of 1-hour on-source time integrations. We discuss the correlation between radio and X-ray luminosities ($L_\mathrm{6 cm}/L_\mathrm{2-10keV}$) in our sub-parsec scale study of higher sensitivity radio observations. Additionally, we explore radio-loudness as a function of a source's accretion rate and the FP proposed for the highly accreting black holes. To conclude, we discuss a physically motivated model to describe the non-detections as a result of the self-synchrotron absorption of particles accelerated in shocks or outflows. 
The sample selection, reliable measurements of X-ray data, higher angular resolution VLBA radio observations, and data calibration details are discussed in Section~\ref{sec:Methodology}, and we present our results and discussion in understanding the origin of the radio emission for this FRAMEx sample comprised of radio-quiet AGNs in Section~\ref{sec:res} and Section~\ref{sec:dis}, respectively.

\section{Methodology} \label{sec:Methodology}
\subsection{Sample Selection} \label{subsection: Sample Selection}

\begin{deluxetable*}{lrrlcccHHH} 
\caption{Observation Source List}
\tablehead{\colhead{Target} & \colhead{R.A.\ (ICRS)} & \colhead{Decl.\ (ICRS)}  & \colhead{Type} & \colhead{Redshift} & \colhead{Distance} & \colhead{log($M_\mathrm{BH}$)} \\
[-0.3cm]
& \colhead{(deg)}  & \colhead{(deg)}    &  & & \colhead{(Mpc)} & \colhead{[$M_{\sun}$]}}
\startdata
\hline
New Snapshots$^\star$ \\
\hline
MCG-05-23-016 & 146.91720558 & $-$30.94884951   &  Sy1.9 & 0.0085 & 36.6 & 7.65 & 39.71    &  & \\
NGC 2273  & 102.53602642 & 60.84582513     &  Sy2   & 0.0061 & 26.2 & 7.99 & $\ldots$ &  & \\
NGC 3147  & 154.22355150 & 73.40075317     &  Sy2   & 0.0093 & 40.1 & 8.81 & $\ldots$ &  & \\
NGC 3516  & 166.69775929 & 72.56867643      &  Sy1.2 & 0.0088 & 37.9 & 7.39 & 39.63    &  & \\
NGC 4102  & 181.59597242 & 52.71100601      & Sy2    & 0.0028 & 12.0 & 7.84 & 39.75    &  & \\
NGC 4138  & 182.37416229 & 43.68524117      &  Sy2   & 0.0030 & 12.9 & 7.71 & 38.97    &  & \\
NGC 5728  & 220.59945800 & $-$17.25306162   & Sy1.9  & 0.0093 & 40.1 & 8.25 & 39.68    &  & \\
NGC 7172  & 330.50781558 & $-$31.86965392   & Sy2    & 0.0087 & 37.5 & 8.15 & 39.52    &  & \\
UGC 6728  & 176.31634979 & 79.68154046      & Sy1.2  & 0.0065 & 28.0 & 5.79 & 39.84    &  & \\
\hline
Deep Integrations$^\dagger$ &&&&&&\\
\hline
NGC 1320 & 51.2028681 & $-$3.04226840   & Sy2   & 0.0089 & 38.4 & 7.96 & 40.49 & 0.24 & $ 4.9\times10^{-3}$\\
NGC 2782 & 138.5212787 & $+$40.11369022   & Sy2   & 0.0085 & 36.6 & 6.07 & 41.51 & 2.62 & $ 5.2\times10^{-2}$\\
NGC 3081 & 149.8731005 & $-$22.82631476   & Sy2   & 0.0080 & 34.5 & 7.74 & 40.97 & 0.74 & $ 1.5\times10^{-2}$\\
NGC 3089 & 149.9028701 & $-$28.33129443   & Sy2?  & 0.0090 & 38.8 & 6.55 & \nodata & \nodata &  ~~\nodata \\
NGC 4388 & 186.4449188 & $+$12.66215153   & Sy2   & 0.0084 & 36.2 & 6.94 & 41.57 & 2.93 & 5.9$\times10^{-2}$\\
NGC 4593 & 189.9143400 & $-$5.34417010   & Sy1   & 0.0090 & 38.8 & 6.88 & 40.65 & 0.35 & 7.0$\times10^{-3}$\\
NGC 6814 & 295.6690092 & $-$10.32345792   & Sy1   & 0.0052 & 22.4 & 7.04 & 39.35 & 0.02 & 3.5$\times10^{-4}$\\
NGC 7314 & 338.9424567 & $-$26.05043820   & Sy1.9 & 0.0048 & 20.6 & 6.76 & 39.68 & 0.04 & 7.5$\times10^{-4}$\\
NGC 7465 & 345.5039963 & $+$15.96477472   & Sy2   & 0.0066 & 28.4 & 6.54 & 40.44 & 0.22 & 4.3$\times10^{-3}$\\
\enddata
\tablecomments{$^\star$Observation information for the additional 9 targets observed after Paper I \citep{Fischer_2021ApJ...906...88F}.\\
$^\dagger$Subset of sample chosen from Paper I for the follow up VLBA 4 hour on source integration time observation.}
\label{tab:sample}
\end{deluxetable*}

\begin{figure}
\includegraphics[width=\columnwidth]{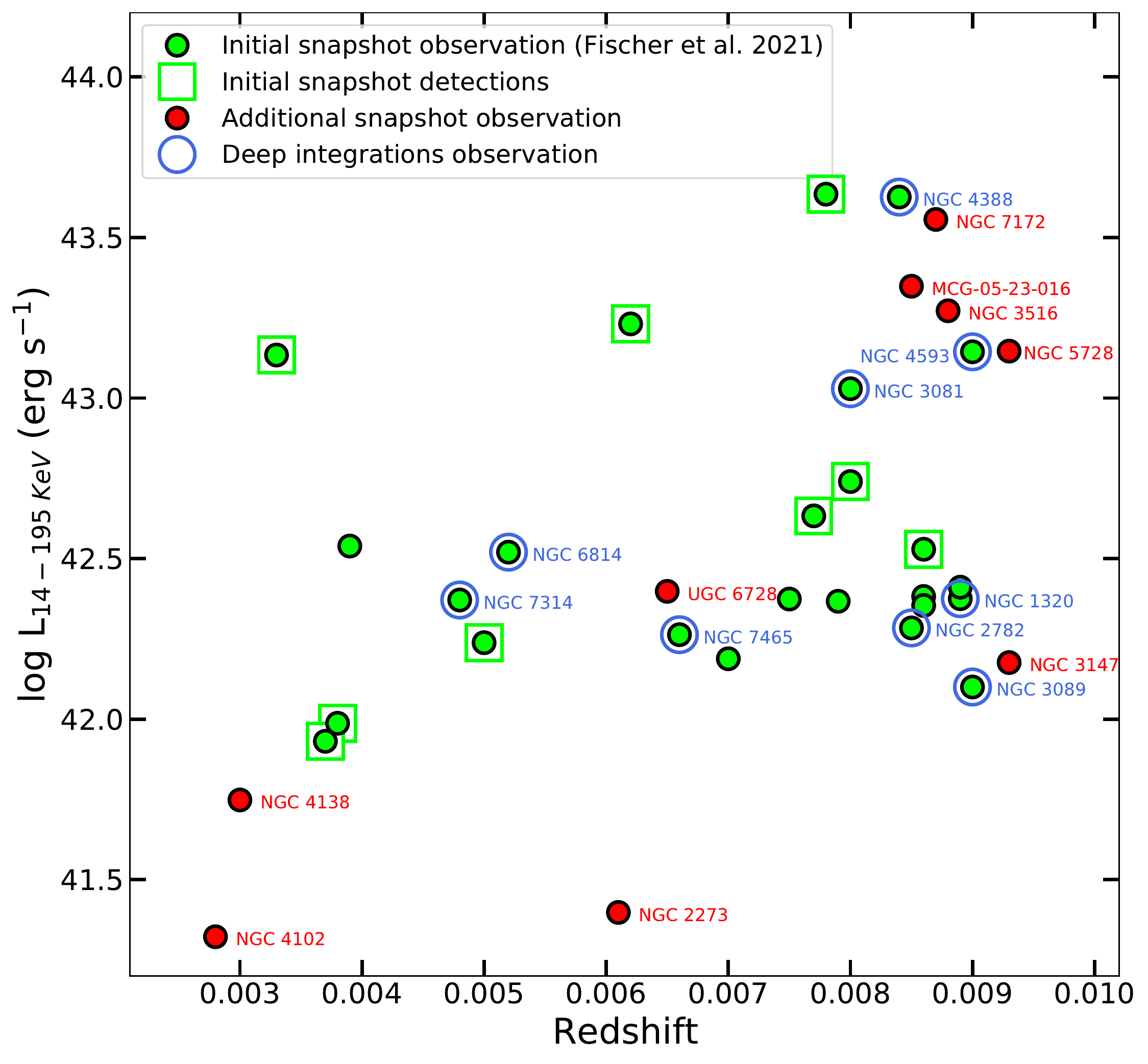}
\caption{Redshift vs hard X-ray (14$-$195 keV) luminosity for AGNs from the 105-month \emph{Swift} BAT catalog. Green filled circles are the sources observed in our initial snapshot observation from Paper~I and green squares denote the detections. The red filled circles are the additional snapshot targets and blue rings identify the 9 deep integration sources.}
\label{fig:detected_sample}
\end{figure}

Since the publication of Paper~I, we have added 9 AGNs to our sample, each observed with the VLBA for 1 hour of on-source integration following the same observing setup as described in Paper~I.  These objects are comprised of four sources above $+60\arcdeg$ (to $\sim +80\arcdeg$) declination, as the original $+60\arcdeg$ limit was based on an initial intent to monitor these objects with The United Kingdom Infra-Red Telescope (UKIRT), which was ultimately not used in the study, plus five sources that are now found to fall within the 40 Mpc distance limit owing to improved redshifts.  These 9 sources are hereafter called ``additional snapshot" observations. We proposed for Target of Opportunity (ToO) time to obtain simultaneous Neil Gehrels Swift Observatory X-ray Telescope (\emph{Swift} XRT) observations for these 9 sources in the same way as we did in Paper~I. 
Alongside the new observations of these additional sources, we re-observed 9 of the 16 non-detected AGNs from Paper~I using the VLBA, with deeper 4-hour on-source integration times 
(see Table \ref{tab:sample}). In order to maximize the likelihood of recovering emission on the milli-arcsecond angular scales of the VLBA we used the archival hundred-parsec-scale C-band, A-array VLA radio structure maps  \citep[see Fig.~3 in][]{Fischer_2021ApJ...906...88F} to select the most point like AGNs for this deeper observation campaign. We illustrate our selection method in Figure~\ref{fig:detected_sample}, and sample sources along with their global properties are listed in Table \ref{tab:sample}. 

\subsection{X-ray Data}

\begin{deluxetable*}{lcc|cc|ccH} 
\setlength{\tabcolsep}{4pt}
\caption{X-ray Data}
\tablehead{ \colhead{~} & \colhead{BAT} & \colhead{~} \vline & \colhead{XRT} & \colhead{~} \vline &   \colhead{NuSTAR$^\dagger$}& & \colhead{~}\\
[-0.1cm]
\hline
\colhead{Target~~~~~~~~~} &  \colhead{F$_{14-195~keV}$} & \colhead{L$_{14-195~keV}$} \vline &  \colhead{F$_{2-10~keV}$} & \colhead{L$_{2-10~keV}$} \vline & \colhead{F$_{2-10~keV}$} & \colhead{L$_{2-10~keV}$}  & \\
[-0.1cm]
\colhead{~} & \colhead{$\times~10^{-11}$} & \colhead{$\times~10^{42}$} \vline& \colhead{$\times~10^{-11}$} & \colhead{$\times~10^{42}$} \vline& \colhead{$\times~10^{-11}$} & \colhead{$\times~10^{42}$} &  \\
[-0.1cm]
\colhead{~} & \colhead{(erg s$^{-1}$ cm$^{-2}$)} & \colhead{(erg s$^{-1}$)} \vline& \colhead{(erg s$^{-1}$ cm$^{-2}$)} & \colhead{(erg s$^{-1}$)} \vline& \colhead{(erg s$^{-1}$ cm$^{-2}$)} & \colhead{(erg s$^{-1}$)} &  }
\startdata
\hline
New Snapshots & & & & & & & \\
\hline
MCG-05-23-016 & $13.9^{+0.30}_{-0.30}$& $22.3^{+0.50}_{-0.50}$ & $7.70^{+0.50}_{-0.50}$~~~ & $12.34^{+0.80}_{-0.80}$ & $9.54\pm 0.02$ & $15.3\pm0.03$ & $13.5\pm0.5$ \\
NGC 2273 & $0.30^{+0.20}_{-0.20}$& $0.25^{+0.20}_{-0.20}$ & $0.020^{+0.007}_{-0.005}\,^{**}$ & $0.016^{+0.006}_{-0.004}$ & $3.60\pm1.30$ & $32.0\pm11.0$ & $19.5\pm5.8$ \\
NGC 3147 & $0.80^{+0.30}_{-0.30}$& $1.50^{+0.60}_{-0.60}$ & $0.14^{+0.07}_{-0.05}$~~~ & $0.27^{+0.13}_{-0.10}$ & $0.28\pm0.04$ & $0.55\pm0.01$ & \nodata~~  \\
NGC 3516 & $10.9^{+0.30}_{-0.30}$& $18.7^{+0.50}_{-0.50}$ & $3.60^{+0.30}_{-0.30}$~~~ & $6.19^{+0.52}_{-0.52}$ & $0.60\pm0.05$ & $1.00\pm0.08$ & \nodata~~  \\
NGC 4102 &  $1.20^{+0.10}_{-0.10}$& $0.21^{+0.02}_{-0.02}$ & $1.70^{+0.80}_{-0.60}\,^{**}$ & $0.29^{+0.14}_{-0.10}$ & $0.92\pm0.03$ & $0.16\pm0.01$ & \nodata~~  \\
NGC 4138 & $2.80^{+0.40}_{-0.40}$ &$0.56^{+0.08}_{-0.08}$  & \nodata~~ & \nodata & \nodata & \nodata &\nodata~~  \\
NGC 5728 & $7.00^{+3.00}_{-2.00}$& $14.0^{+6.00}_{-4.00}$ & $5.00^{+3.00}_{-3.00}\,^{**}$ & $9.62^{+5.77}_{-5.77}$ & $4.12\pm0.08$ & $7.95\pm0.16$ & $8.3\pm2.9$ \\
NGC 7172 & $21.4^{+0.50}_{-0.50}$& $36.0^{+0.80}_{-0.80}$ & $0.40^{+0.30}_{-0.20}$~~~ & $0.67^{+0.50}_{-0.34}$ & $8.38\pm0.08$ & $14.0\pm0.10$ & $10.3\pm3.5$ \\
UGC 6728 & $2.70^{+0.20}_{-0.20}$ & $2.50^{+0.20}_{-0.20}$ & $0.60^{+0.10}_{-0.10}$~~~ & $0.56^{+0.09}_{-0.09}$ & $1.44\pm0.02$ & $1.35\pm0.02$ & $1.5\pm0.5$ \\
\hline
Deep Integrations$^{**}$ & & & & & & & \\
\hline
NGC 1320 & $1.34^{+0.04}_{-0.07}$ & $2.36^{+0.07}_{-0.12}$ & \nodata~~ & \nodata & $0.54\pm0.05$ & $0.95\pm0.09$&  $1.0\pm0.4$\\
NGC 2782 & $1.20^{+0.04}_{-0.07}$ & $1.92^{+0.06}_{-0.11}$  & $0.13^{+0.03}_{-0.02}$ & $0.21^{+0.05}_{-0.03}$ & \nodata & \nodata& \nodata~~\\
NGC 3081 & $7.50^{+0.04}_{-0.08}$ & $10.68^{+0.06}_{-0.011}$ & $3.16^{+0.14}_{-0.27}$ & $4.50^{+0.20}_{-0.38}$ & $6.57\pm0.66$ &$9.26\pm0.09$&   $9.5\pm3.6$\\
NGC 3089 & $0.70^{+0.02}_{-0.08}$ & $1.26^{+0.04}_{-0.14}$ & $0.20^{+0.02}_{-0.01}$ & $0.36^{+0.04}_{-0.02}$ & \nodata & \nodata & \nodata~~\\
NGC 4388 & $26.98^{+0.06}_{-0.06}$& $42.31^{+0.09}_{-0.09}$ & $2.60^{+0.80}_{-0.60}$ & $4.10^{+1.00}_{-0.90}$ &  $1.39\pm0.06$ &$2.17\pm0.09$&  $4.3\pm1.6$\\
NGC 4593 & $7.73^{+0.11}_{-0.11}$ & $13.93^{+0.20}_{-0.20}$ & $3.55^{+0.05}_{-0.05}$ & $6.40^{+0.09}_{-0.09}$ & $2.28\pm0.02$ &$4.10\pm0.04$&  $3.9\pm1.3$\\
NGC 6814 & $5.52^{+0.12}_{-0.03}$ & $3.31^{+0.07}_{-0.02}$ & $2.09^{+0.03}_{-0.03}$ & $1.26^{+0.02}_{-0.02}$ & $3.48\pm0.03$ &$2.07\pm0.02$&  $1.8\pm0.6$\\
NGC 7314 & $4.62^{+0.09}_{-0.04}$ & $2.35^{+0.05}_{-0.02}$ & $2.04^{+0.03}_{-0.03}$ & $1.04^{+0.02}_{-0.02}$ & $3.80\pm0.04$ &$1.93\pm0.02$&  $1.3\pm0.5$\\
NGC 7465 & $1.90^{+0.05}_{-0.07}$ & $1.83^{+0.05}_{-0.07}$ & $1.12^{+0.02}_{-0.02}$ & $1.08^{+0.02}_{-0.02}$ & $1.26\pm0.02$  &$1.22\pm0.02$& $1.2\pm0.4$\\
\enddata
\tablecomments{$^{**}$Archive fluxes with inflated errors based on 105-month \emph{Swift} BAT variability (see Section \ref{subsection: XRT} for details).\\
$^\dagger$Nustar X-ray spectra results for individual sources is added in Appendix~\ref{appendix:a}. }
\label{tab:xray_data}
\end{deluxetable*}

\subsubsection{\emph{Swift} BAT Data}

We use archival BAT data from the \emph{Swift} BAT 105-month Hard X-ray Survey \citep{Oh_2018ApJS..235....4O}, to compare the long term luminosity for each object. When solely examining the BAT data, $N_{\rm H}$ is difficult to constrain except in the case for Compton thick sources at this energy range. Therefore, we fit the BAT data with archival \nustar\ data. This improves the overall fit for each object and provided a more accurate normalization. We fit the BAT spectrum using a similar method as described in Paper~I using a simple power law model. However, for some of the targets this did not suffice and they needed an additional reflective component that allowed the cutoff energy to vary. Therefore, we utilized the model \texttt{pexrav} to accomplish this.  See Table \ref{tab:xray_data} for resulting fluxes and luminosities.

\subsubsection{\emph{Swift} XRT Data}\label{subsection: XRT}
We obtained \emph{Swift} XRT time using the ToO proposal mechanism for the additional VLBA snapshot sources but we did not propose for simultaneous \emph{Swift} XRT observations for the deeper integration time objects shown in Table \ref{tab:sample}, and 
not all of the additional snapshot targets were observed with ToO due to competing schedule priorities. For the unobserved targets, archival XRT data is used. Analysis followed the same procedure as described in Section 2.6 of Paper~I.

Unlike in Paper~I, for some of the additional snapshot targets and for all of the deep integration objects, the archival XRT data are not contemporaneous with the radio observations, so we add an additional uncertainty term in quadrature to the formal flux uncertainties, corresponding to each source variability. To estimate this, we use the source X-ray light curves from the BAT 105-month catalog, and for each source determine the intrinsic scatter term such that the reduced $\chi^2$ of the source light curve is unity. This is typically about $\sim0.1-0.2$~dex. 
Table \ref{tab:xray_data} lists the X-ray fluxes and luminosities achieved from the spectral fits.

\subsubsection{NuSTAR Data}\label{subsubsec:NuSTAR}
Determining the spectral properties of the X-ray emitting corona in AGNs is of crucial importance to shed light on their central engine and on the link between accretion and ejection phenomena. In recent years, an important role has been played by the Nuclear Spectroscopic Telescope Array (\nustar), a focusing hard X-ray telescope launched in 2012 with a large effective area and excellent sensitivity in the energy range of 3--80 keV, which makes it possible to tighly constrain the contributions of absorption and reflection and hence measure the intrinsic properties of the primary X-ray emission. The goal of our \nustar\ data analysis was 
to use the column densities, which are generally less variable, and to determine if XRT is seeing the intrinsic X-ray continuum or heavily absorbed or scattered continuum. In addition, we compared the \nustar\ data with Swift XRT data to assess whether the X-ray luminosities might be a source of FP discrepancy. 

Eight out of nine of the additional VLBA snapshot targets and seven out of nine of the deeper VLBA observation targets possess archival \nustar\ data (3--79~keV) from observations carried out recently or reprocessed very recently and therefore did not need to be reprocessed with the data analysis pipeline \texttt{nupipeline}. From the calibrated event files we extracted light curves and spectra, along with the RMF and ARF files necessary for the  spectral analysis, using the \texttt{nuproduct} script. All spectra were binned with a minimum of 20 counts per bin using the HEASoft task \texttt{grppha} 3.0.1 for the $\chi^2$ statistics to be valid. 
We performed the X-ray spectral analysis using the \xspec\ \texttt{v.12.9.0} software package \citep{arnaud_1996ASPC..101...17A}, and  the errors quoted on the spectral parameters represent the 1$\sigma$ confidence level. Our baseline model, expressed in the \xspec\ syntax , is:
\begin{verbatim}
phabs * (atable(Borus) + MYTZ*BMC)
\end{verbatim}
where the first absorption model \texttt{phabs} accounts for our Galaxy contribution, the \borus\ table model parametrizes the continuum scattering and fluorescent emission line components associated with the torus  \citep{Balokovi__2018}, and \texttt{MYTZ} models the absorption and Compton scattering acting on the transmitted primary emission \citep{Murphy_2009MNRAS.397.1549M}, which is described by the Comptonization model \texttt{BMC} \citep{titarchuk_1997ApJ...487..834T}. For objects not classified as Seyfert 2, the \texttt{MYTZ} component is substituted by a \texttt{zphabs} model left free to vary. Depending on the complexity of the X-ray spectra, additional components (such as individual lines, additional absorption and scattering components, or the fraction of primary emission directly scattered towards the observer by a putative optically thin ionized medium) may be included and described in the text for individual sources (See Appendix~\ref{appendix:a}).

\subsection{VLBA Data}  \label{subsection: VLBA}

\subsubsection{Observations}


\begin{figure*}[htp]
\centering
{\color{red}
\stackinset{l}{1.65in}{b}{1.2in}{\rotatebox{21}{\rule{2.15in}{1pt}}}{
\stackinset{l}{1.61in}{b}{0.255in}{\rotatebox{-25.4}{\rule{2.2in}{1pt}}}{
\includegraphics[width=.4\textwidth]{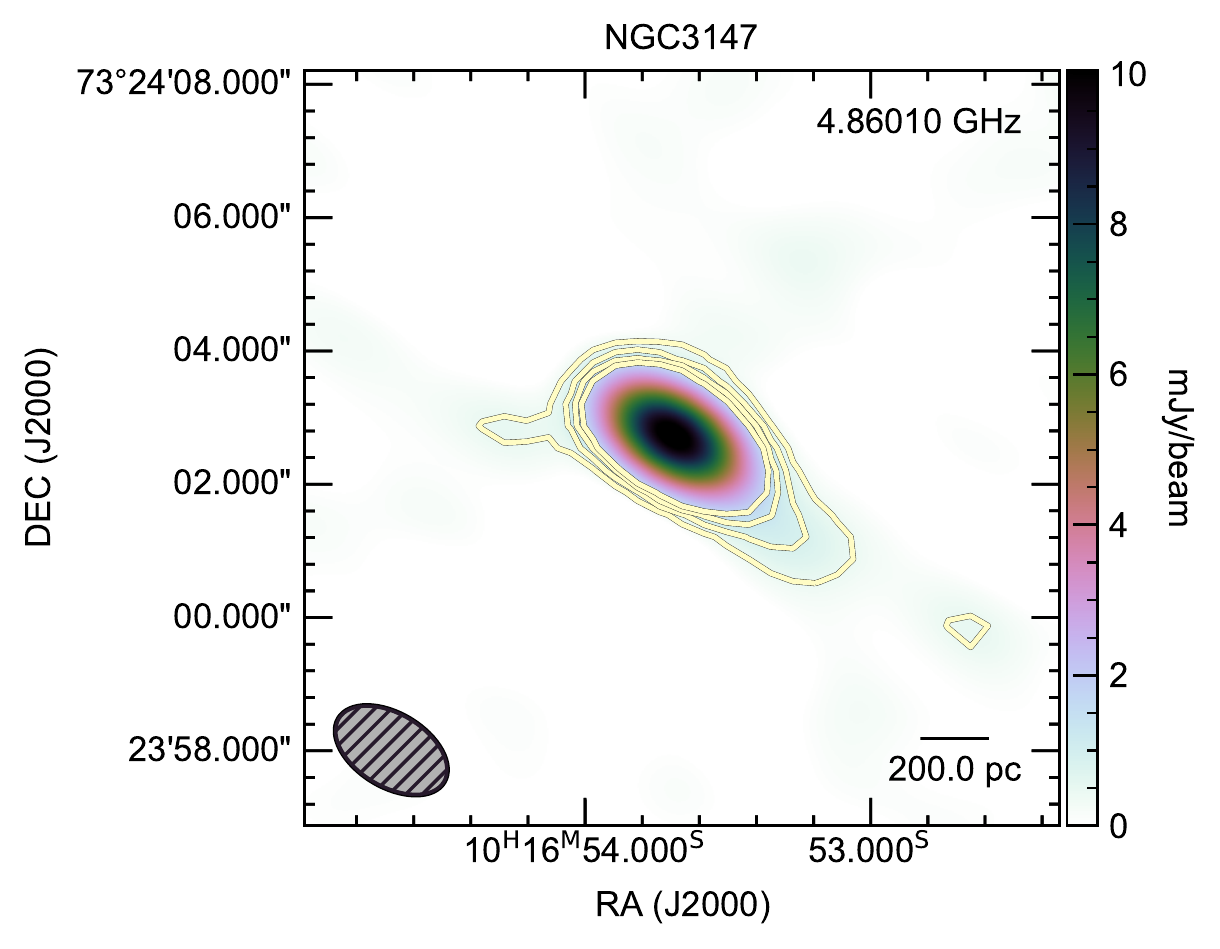}
\fcolorbox{red}{white}{\includegraphics[width=.4\textwidth]{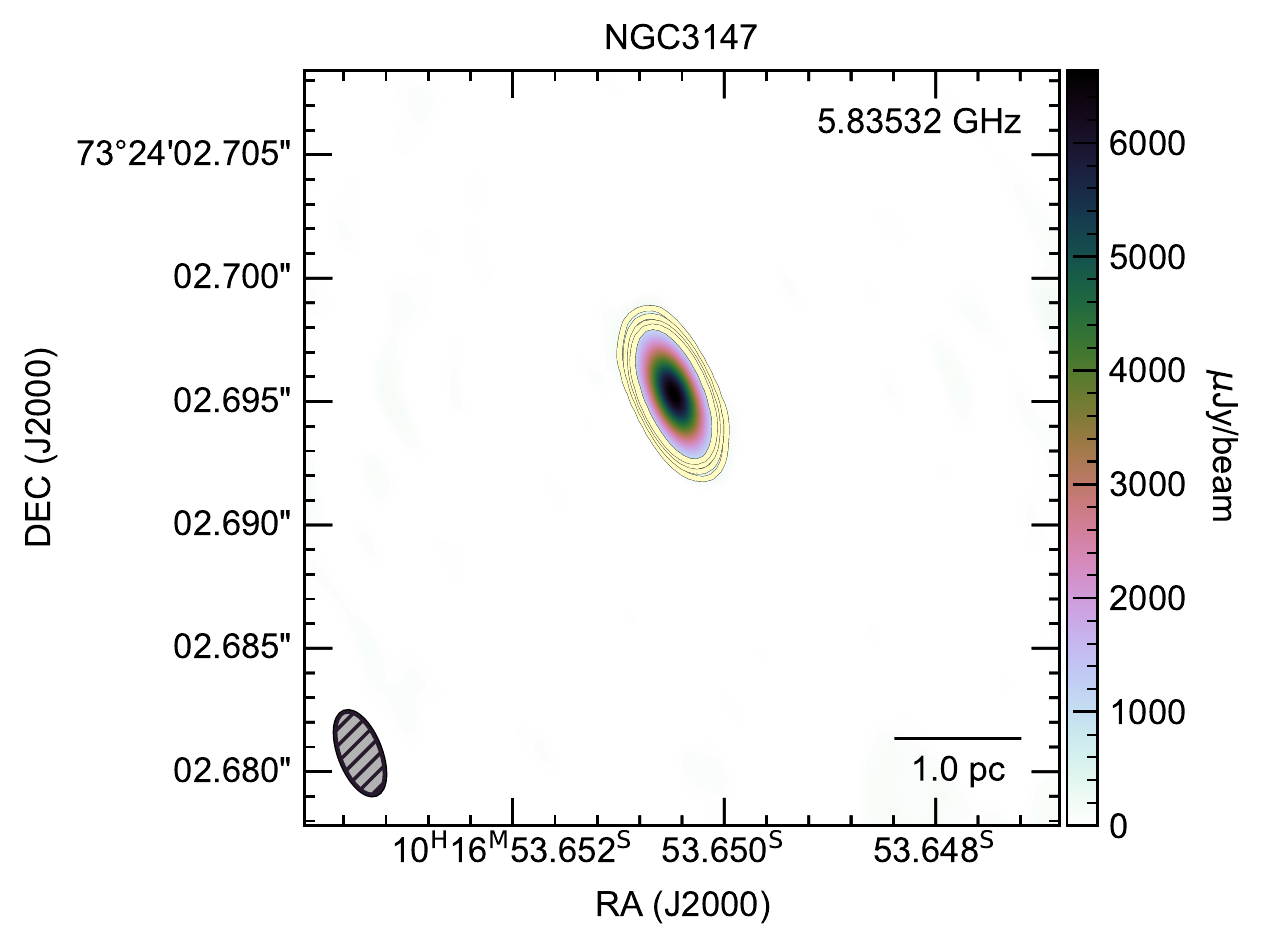}}}}}
\fcolorbox{red}{white}{\includegraphics[width=.405\textwidth]{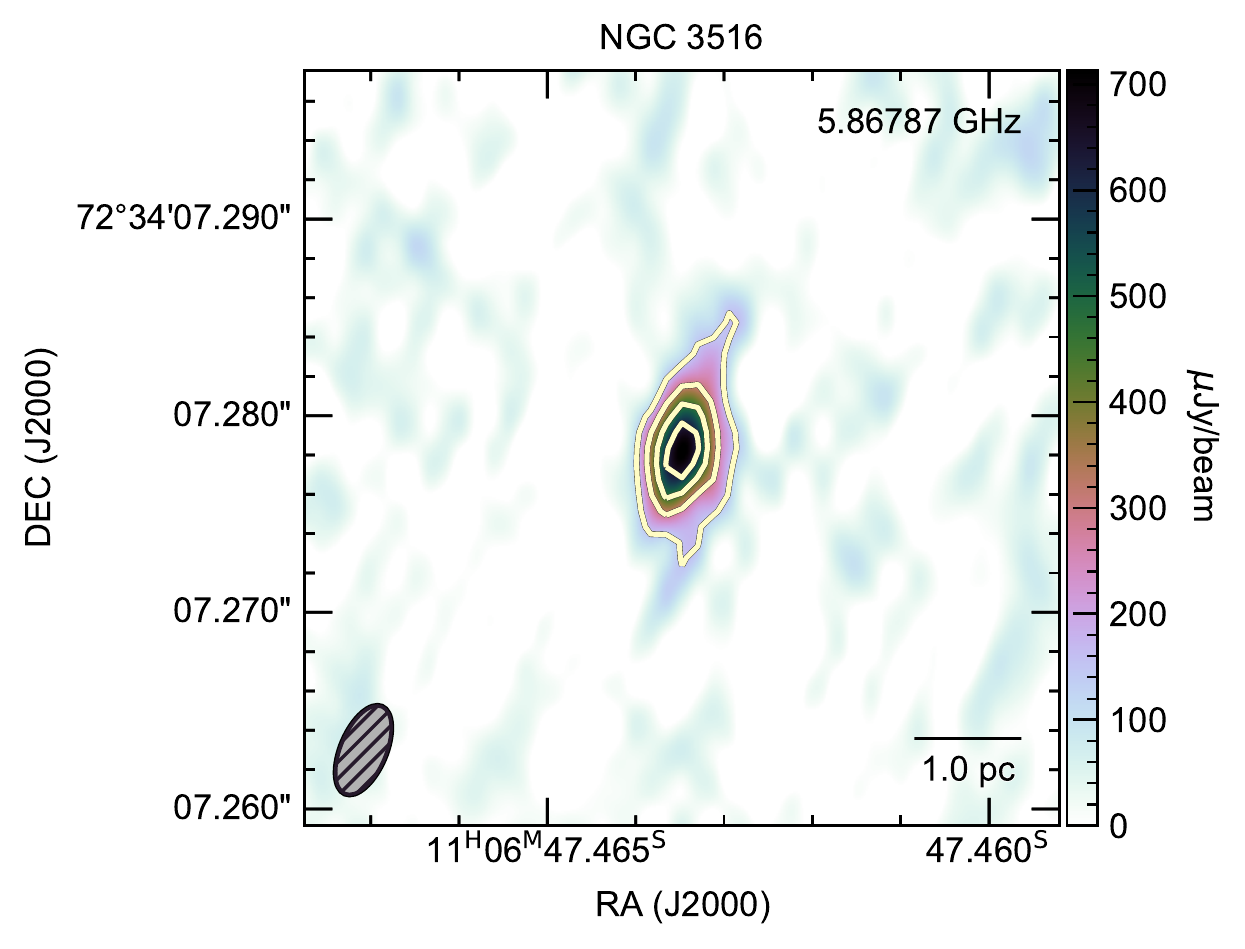}}
\fcolorbox{blue!70}{white}{\includegraphics[width=.4\textwidth]{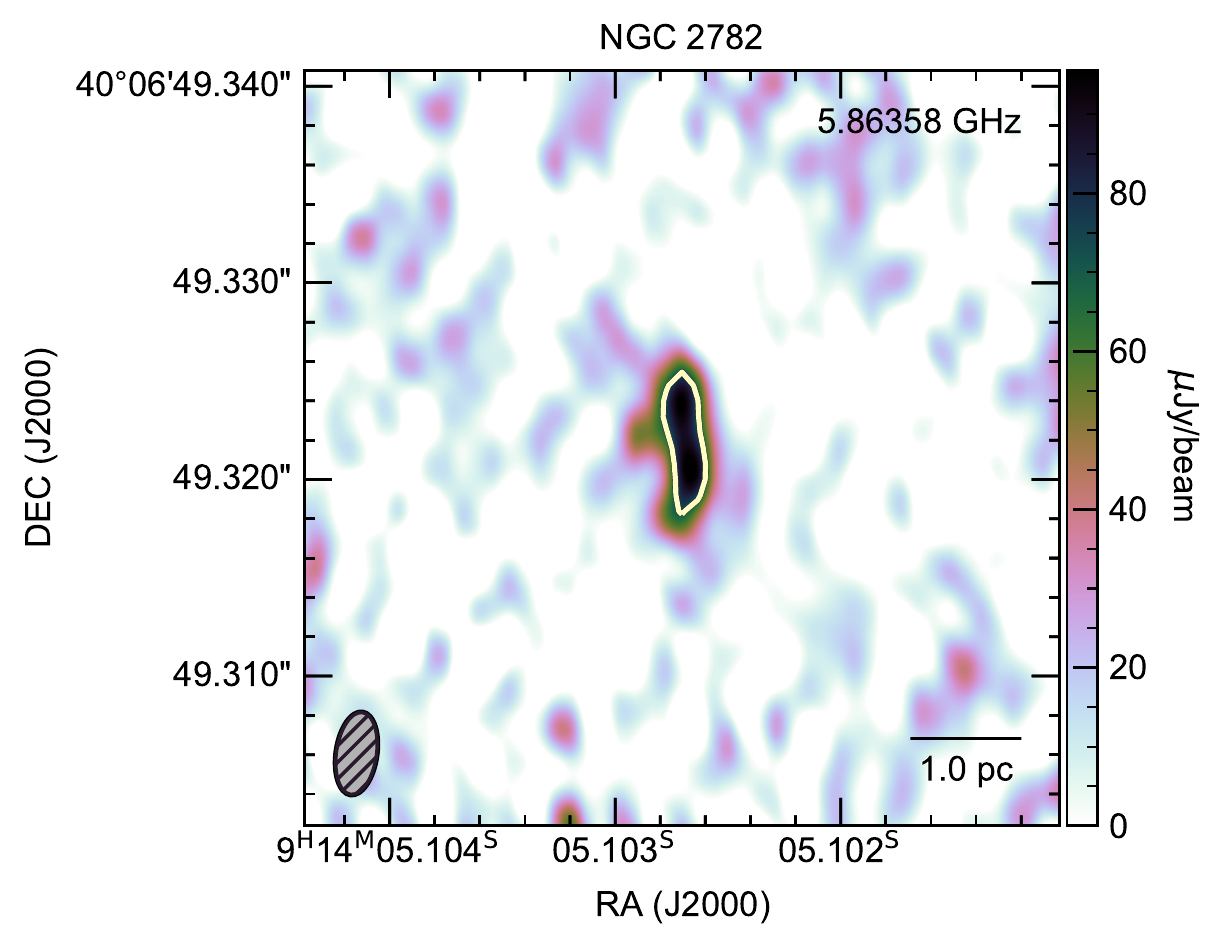}}
\fcolorbox{blue!70}{white}{\includegraphics[width=.405\textwidth]{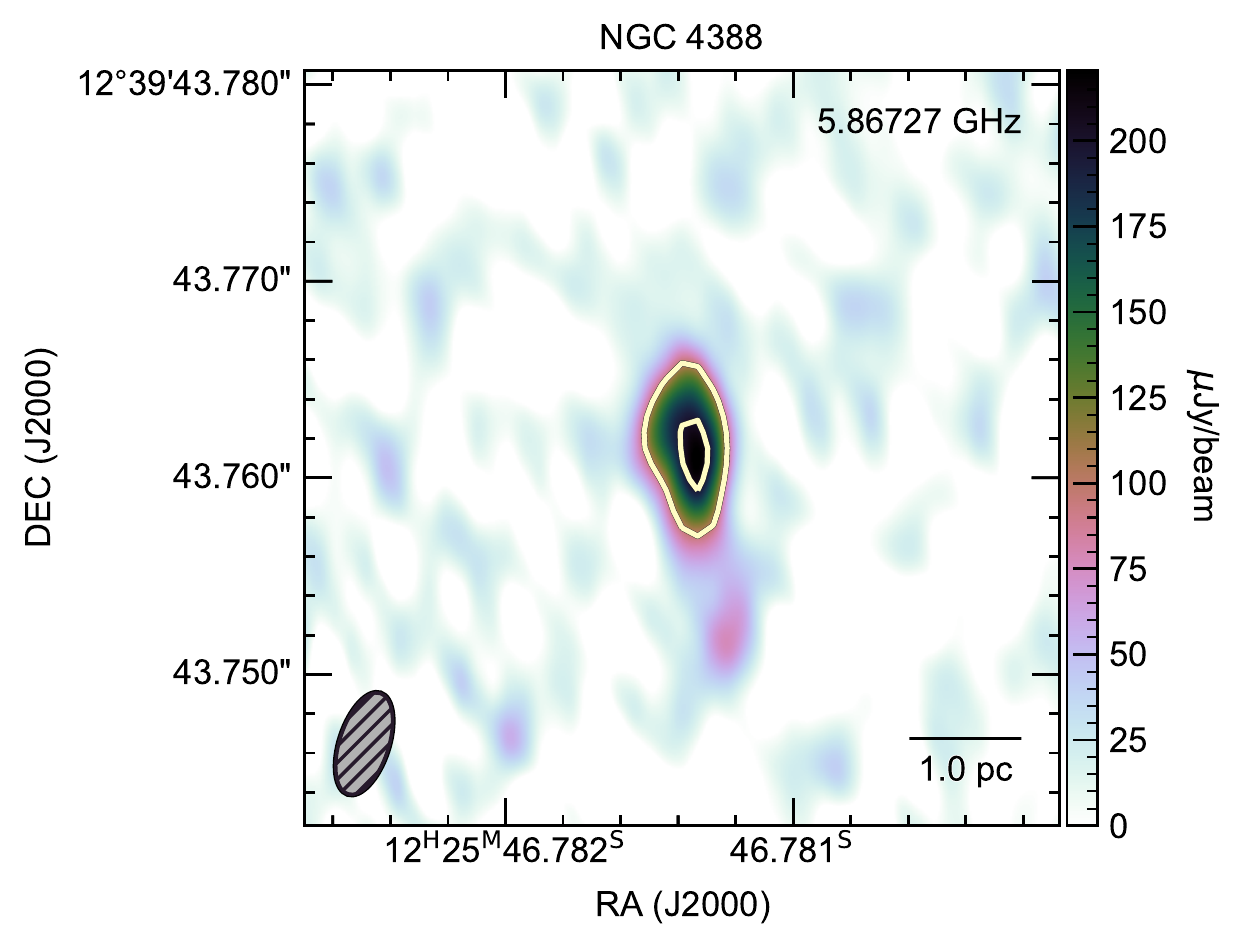}}
\fcolorbox{blue!70}{white}{\includegraphics[width=.40\textwidth]{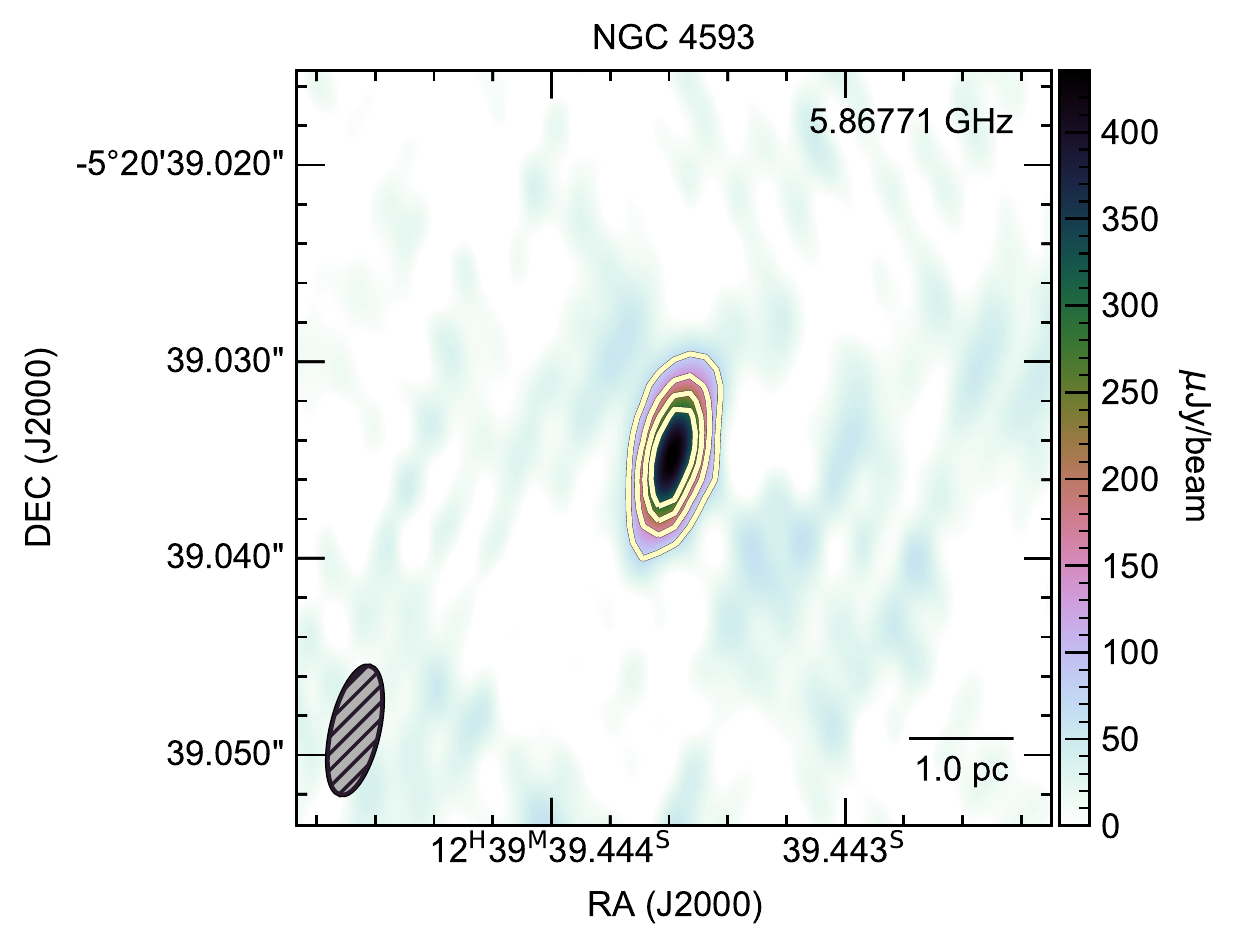}}
\caption{C-band (5.8~GHz) parsec-scale radio detections for the VLBA additional snapshot observations ({\it red boxes}) and for the VLBA deep integration observed sources ({\it blue boxes}). The outermost contour represents the 5$\sigma$ flux limit above RMS, and the interior contours increase as $\sigma \times$(10, 15, 20). The gray ellipses to the lower left of each frame represent the synthesized beam 
size for that observation.~The top left image shows the C-band (4.9~GHz) sub-kiloparsec-scale radio morphology of NGC 3147 from the VLA archive, demonstrating the huge difference in spatial scales and structures between VLA and VLBA observations.}
\label{fig:radio_det_sources}
\end{figure*}


\begin{deluxetable*}{llcrccHH} 
\label{tab:radio_obs}
\tablecaption{VLBA Observations}
\tablehead{\colhead{Target} & \colhead{Frequency} & \colhead{Restoring Beam}                 & \colhead{Beam angle} & \colhead{RMS} & \colhead{Calibrator}\\
& \colhead{(GHz)}       &  \colhead{($\alpha \times \delta$; mas)}	& \colhead{(deg)} & \colhead{($\mu$Jy bm$^{-1}$)}  & \colhead{IERS Name}}
\startdata
\hline
New Snapshots$^{*}$ \\
\hline
MCG-05-23-016 & 5.86747 & 9.37$\times$3.76 & $-$3.76 & 32.6 & J0948$-$2901 &  \\
NGC 2273 & 5.84801 & 3.66$\times$1.58 &$-$23.6 & 31.2 & J0638+5933 & 102.536026417 & 60.8458251306 \\
NGC 3147 & 5.83532 & 3.53$\times$1.61 & 21.2 & 46.0 & J1027+7428 & 154.223551500 & 73.4007531694 \\
NGC 3516 & 5.86788 & 4.80$\times$2.29 & $-$23.0 & 29.5 & J1048+7143 & \\
NGC 4102 & 5.86787 & 3.20$\times$1.96 & $-$24.3 & 16.9 & J1200+5300 & 181.595972417 & 52.7110060111 \\
NGC 4138 & 5.86740 &4.43$\times$2.19 & $-$8.3 & 24.9 & J1221+4411 &  \\
NGC 5728 & 5.86568 & 7.29$\times$3.11 & $-$9.2 & 28.1 & J1445$-$1629 & 220.599458000 & -17.2530616194 \\
NGC 7172 & 5.86068 & 3.61$\times$2.91 & 11.1 & 35.3 & J2158$-$3013 & 330.507815583 & -31.8696539194 \\
UGC 6728 & 5.86786 & $3.98\times1.86$ & $-$35.3 & 22.6 & J1058+8114 & 176.316349792 & 79.6815404611 \\
\hline
Deep Integrations &&&&&&\\
\hline
NGC 1320 & 5.86337 & 7.47$\times$2.02 &$-$16.1~~ &  16.2~~ & J0321$-$0526  & 50.499459824(8.17) & $-$5.4367857(222.2)\\
NGC 2782 & 5.86358 & 4.20$\times$2.04 &  $-$8.5~~ &  13.7~~ & J0916+3854  & 139.203769060(7.04) &  38.9078184(115.6) \\
NGC 3081 & 5.86592 & 6.94$\times$3.40 & $-$2.5~~ &  16.7~~ & J1006$-$2159  & 151.693390348(4.80) &$-$21.9890028(109.5)\\
NGC 3089 & 5.87245 & 8.27$\times$3.82 &   4.8~~ &  20.4~~ & J1011$-$2847  & 152.772989020(7.74) &$-$28.7945604(246.8)\\
NGC 4388 & 5.86727 & 5.37$\times$2.49 &  $-$18.1~~ &  19.4~~ & J1225+1253  & 186.265597247(4.1) &	12.8869831(138)\\
NGC 4593 & 5.86771 & 6.66$\times$2.37 & $-$12.9~~ &  14.7~~ & J1248$-$0632  & 192.095731923(4.98) &	$-$6.5360605(154.7)\\
NGC 6814 & 5.80882 & 7.55$\times$3.13 &$-$8.5~~ &  27.1~~ & J1939$-$1002  & 294.988569046(4.17) &$-$10.04486683(98.1)\\
NGC 7314 & 5.80288 & 9.02$\times$2.94 &$-$8.3~~ &  27.8~~ & J2243$-$2544  & 340.860036591(6.44) &$-$25.7418576(189.9)\\
NGC 7465 & 5.80877 & 2.93$\times$1.40 &$-$5.5~~ & 43.0~~ & J2300+1655  & 345.179129648(7.01) &	16.9206644(148.4)\\
\enddata
\tablecomments{$^{*}$Observed in September 2020.
}
\end{deluxetable*}

Using the EVN Calculator, four hours of on-source integration time with the VLBA at 6 cm (C-band) produces a theoretical thermal noise rms of $\sim$10 $\mu$Jy/beam, more sensitive than the observations in our initial snapshot program by a factor of two, allowing for potential 10-sigma or higher detections for the proposed 9 deep integration targets. These new observations have $uv$-coverage similar to our initial snapshot program which interleaved observations of multiple targets, but now by sampling the $uv$-plane more finely over fuller, 4-hour long on-source integrations per target.
We used phase referencing with known nearby calibrator sources for accurate phase calibration, which required total telescope times of roughly 6$-$6.5 hours per schedule per source.  This resulted in extending the sampling of parallactic angles and $uv$-coverage for high fidelity imaging. The estimated absolute flux density calibration is within 5-10\%, which is the nominal VLBA flux calibration uncertainty.  


\subsubsection{Calibration and Imaging}
We used National Radio Astronomy Observatory (NRAO) Astronomical Image Processing System \citep[\textsc{aips};][]{Van_1996ASPC..101...37V} release 31DEC19 to calibrate our VLBA data. Each dataset was calibrated independently by target -- phase calibrator pair.
Bad data were flagged and calibration was performed using the standard \textsc{aips} VLBA procedures, following the prescription outlined in Paper~I. 
We used the \textsc{aips} task {\sc imagr} to make images of the calibrators and sources, cleaning the images until the rms approached the theoretical thermal limit. 
We achieved a S/N ratio of $>$10 for all our detected sources.  For more information on our calibration and imaging procedures, please see Sections 2.2.1 and 2.2.2, respectively, in Paper~I.

\section{Results} \label{sec:res}

\begin{deluxetable*}{l l c c c c l r H H H } 
\tabletypesize{\scriptsize}
\tablecaption{Radio Image Properties of the Detected Sources}
\label{tab:radio_xray_luminosity}
\tablehead{\colhead{Name} & \colhead{F$_{peak}^a$}   & \colhead{Log\,F$_{peak}$} & \colhead{Log\,L$_{peak}$}&  \colhead{S$_{int}^b$} &\colhead{T$_{b}$}& \colhead{Source R.A.\ (ICRS)} & \colhead{Source Decl.\ (ICRS)}\\ 
[-0.2cm]
& \colhead{(mJy bm$^{-1}$)}	 &\colhead{($\times10^{-17}$\,erg\,s$^{-1}$\,cm$^{-2}$)} 	& \colhead{(erg s$^{-1})$} &   \colhead{(mJy)} & (K) & \colhead{(HMS)}  & \colhead{(DMS)}}
\startdata
\hline  
New Snapshots &&&&&&&\\
\hline
NGC 3147      &  ~~~~$6.698\pm0.042$      &39.085$\pm$0.245 &37.86 & $6.228\pm0.077$  & 10$^{7.6}$& 10:16:53.650476$\pm$ 0.000001 & 73:24:02.69529$\pm$0.00001 & 3.43 & 1.54 & 23 \\
NGC 3516      &  ~~~~$0.672\pm0.029$   &3.943$\pm$0.170 & 36.82&  $1.274\pm0.083$ & 10$^{6.3}$ &  11:06:47.46346$\pm$0.00001 & 72:34:07.2783$\pm$0.0002& 6.75 & 3.08 & 170 \\
\hline
Deep Integrations &&&&&&&\\
\hline
NGC 2782       & ~~~~0.098$\pm$0.007      & 0.575$\pm$0.041 & 35.95 &  0.336$\pm$0.033 & 10$^{5.6}$ &09:14:05.10270$\pm$0.00001 & 40:06:49.3224$\pm$0.0005 & 10.93 & 3.25 & 4 \\
NGC 4388     & ~~~~0.214$\pm$0.018      &1.256$\pm$0.105 & 36.28 &  0.563$\pm$0.065 &10$^{5.8}$ & 12:25:46.78136$\pm$0.00001 & 12:39:43.7611$\pm$ 0.0004    & 9.40 & 3.45 & 9 \\
NGC 4593  & ~~~~0.424$\pm$0.023      &2.488$\pm$0.135 & 36.64&  0.498$\pm$0.049   & 10$^{6.0}$ &   12:39:39.443589$\pm$0.000004 & $-$05:20:39.0347$\pm$0.0002 & 6.58 & 2.63 & 168  \\
\enddata
\tablecomments{$^{a}$Peak flux values are derived from CASA's 2-D Gaussian model fitting algorithm and\\
$^{b}$integrated flux densities are determined from 5$\sigma$ outermost contour region shown in Figure~\ref{fig:radio_det_sources}.}
\end{deluxetable*}

In Table \ref{tab:radio_obs}, we list the radio observation properties for our final calibrated images displayed in Figure \ref{fig:radio_det_sources} for the additional snapshot and deep integration samples.
Table~\ref{tab:radio_xray_luminosity} lists the peak flux values of the detections  derived from CASA’s 2-D Gaussian model fitting algorithm and the integrated  flux densities measured using the 5$\sigma$ outermost contour shown in our images. For the additional snapshot observation sample, the nuclei for both the detections (NGC 3147 and NGC 3516) appear unresolved, but with follow up higher sensitivity deep observations, the core for NGC 2782 and NGC 4388 appear slightly elongated. The central component for NGC 4593 remains unresolved, similar to the large scale C-band VLA archival\footnote{\url{http://www.aoc.nrao.edu/~vlbacald/src.shtml}} image shown in Paper I. In the C-band VLA image of NGC 4388, a resolved additional component separated by $\sim$200 pc from central core is seen towards the south-west direction. Our VLBA image also shows a similar elongated feature towards the same direction, but closer to the central peak ($\sim$1 pc).  For NGC 2782, a large kpc-scale wing-like structure primarily spread in the north-south direction in the VLA  A-configuration\footnote{\url{https://public.nrao.edu/vla-configurations/}} C-band image also exhibited a similar elongation in our VLBA deep observation on very small scales (pc, as compared with kpc) around the central core of the AGN. However, NGC 4593, even with the higher sensitivity observation, remains an unresolved point-like source with a luminous radio core contained within.  The region generating the radio emission in its nucleus is extremely small ($\leq$1 pc), and comparable in size to the accretion disk~\citep[][]{Hawkins_2007A&A...462..581H}{}. 


\begin{figure}
\includegraphics[width=\columnwidth]{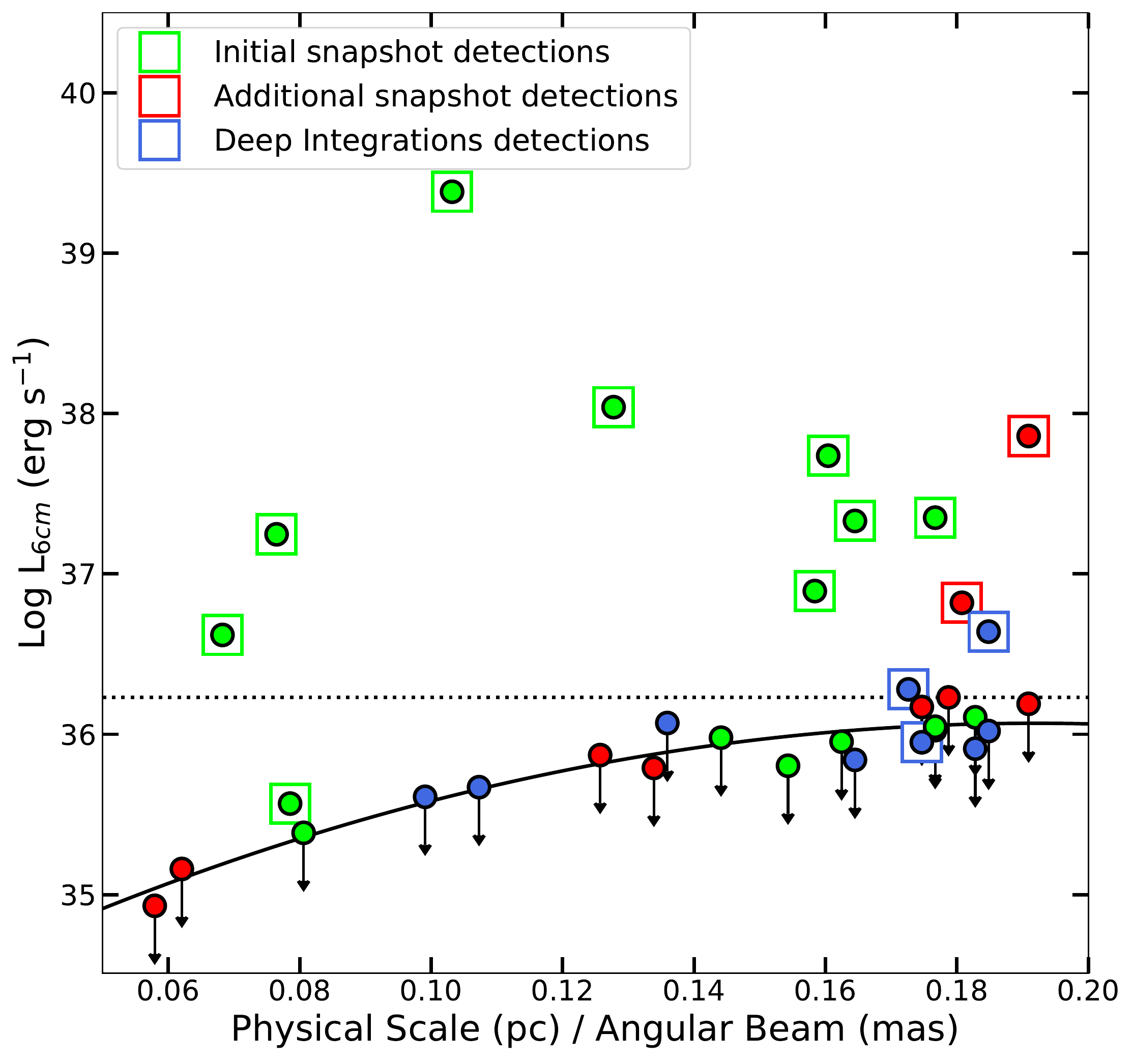}
\caption{Radio luminosity vs the physical scale/beam size (pc mas$^{-1}$) for all the detected and non detected sources (5 $\sigma$ upper limits). The downward arrows are 5 times the 1$\sigma$ radio noise level in each cleaned image, and the black solid line is the fit for all the non-detections. It denotes the average noise level at a given scale derived from our calibrated images. The black dotted line represents the non-detection upper limit across the whole distance range of the sample.}
\label{fig:detected_sample_rms}
\end{figure}

Prior to our data analysis, we looked over our sample again to confirm that the sources are free from any biasing in determining the radio detections within the redshift range (0.028 $<$ z $<$ 0.093) of our sample. Figure \ref{fig:detected_sample_rms} shows the radio luminosity distribution of our total volume-complete sample of 34 AGNs as a function of the ratio between the physical and angular size ($\sim$redshift) out to our 40 Mpc limit. The colored circles enclosed with boxes are the detections and those with the arrows are the 5$\sigma$ upper limits for the non detections. The black solid line is the fit for the average 5$\sigma$ rms for the non-detected sources and demonstrates the limit for any detection in the resulting images. We explored the possibility that some of our distant objects may suffer from ``beam dilution'' due to the lack of exposure time. For a distant point-like source with a small solid angle ($\Omega_{s}$) compared to the telescope's beam solid angle ($\Omega_{s}$ $\propto$ beam size), the measured signal is ``diluted" by the ratio of the solid angles of the source and beam, respectively. If that were true in our case, the detections would be expected to be primarily from the nearby objects with a drop off towards more distant sources. However,  Figure~\ref{fig:detected_sample_rms} shows the luminosities of our observations as a function of the physical scale to beam solid angle thus demonstrating that our observations likely did not suffer beam dilution. In addition, all the detections needed to be higher than the upper limit of 10$^{36.2}$ erg s$^{-1}$ (dotted line) for non-detections, across the whole distance range. Radio luminosities for the detections except for NGC 1068 and NGC 2782 all lie above this  5$\sigma$ rms line. NGC 1068 is one of the nearest sources from the initial snapshot sample (16.3 Mpc), so its proximity combined with the high sensitivity of the data and extended morphology enabled its detection despite its low radio luminosity. NGC 2782 is one of the more distant deep observations, and its radio luminosity is comparable to the maximum upper limit level.  Both are considered to host AGN that are heavily obscured, with NGC 1068 especially so. 

\begin{figure}
\includegraphics[width=\columnwidth]{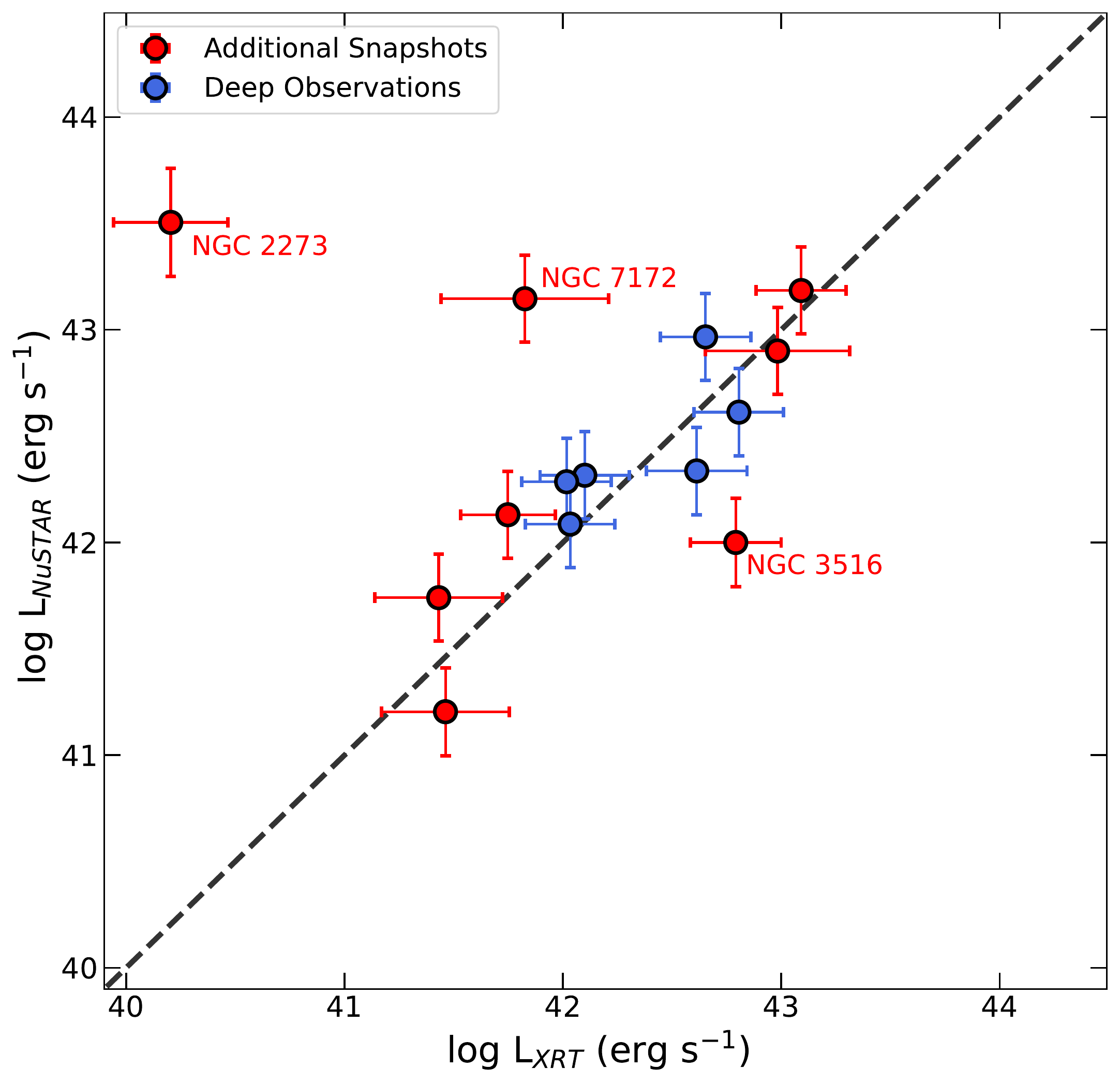}
\caption{Comparison between the soft X-ray (2-10 keV) luminosity of NuSTAR and XRT data from the X-ray scaling method and indirect optical methods, respectively. The black dashed line is the one-to-one line for comparison. Three sources of errors are added in quadrature for each data point as they are all independent: the statistical error of the fit, the error due to inter-epoch variability (previously estimated at 0.1$-$0.2 dex), and the absolute flux calibration error (typically about 10\% for X-ray data).}
\label{fig:soft_xray_comparison}
\end{figure}



Figure~\ref{fig:soft_xray_comparison} compares our XRT and \nustar~X-ray (2-10 keV) luminosities from Table \ref{tab:xray_data} for our additional snapshot (red circles) and deep integration (blue circles) observations. 
NGC 2273, NGC 7172, and to a lesser extent NGC 3516 appear as outliers in  Figure \ref{fig:soft_xray_comparison}, but we note that NGC 2273 is known 
as a Compton-thick AGN and hence its intrinsic X-ray luminosity can not be estimated at the softer X-ray energy range of XRT. 
On the other hand, for NGC 7172, a short-time variability ($\sim$~30\%) in 2-10 keV flux was reported by \citet{Guainazzi_1998MNRAS.298..824G} and a recent study \citep{Mehdipur_2022ApJ...925...84M} explained the new (since 2017) low-flux state and X-ray spectrum variability of NGC 3516 in detail. With the exception of the three sources discussed above, the 2-10 keV luminosities obtained from the \nustar\ are consistent with the \emph{Swift} XRT measurements within their respective uncertainties (see section \ref{subsection: XRT}). Although the \nustar\ observations are not simultaneous with the radio ones, which were taken at the same time as the \emph{Swift} XRT pointings, the good correlation between \nustar\ and XRT values indicates that the X-ray variability does not play a significant role for our sample. We used \nustar\ flux data (reported from Table \ref{tab:xray_data}) to properly account for the effects of absorption and reflection, and all the individual \nustar\ spectral analysis for the sources are shown in Appendix~\ref{appendix:a}.

\section{Discussion} \label{sec:dis}

\subsection{Radio$-$X-ray Correlation}

\begin{figure}
\centering
\includegraphics[width=\columnwidth]{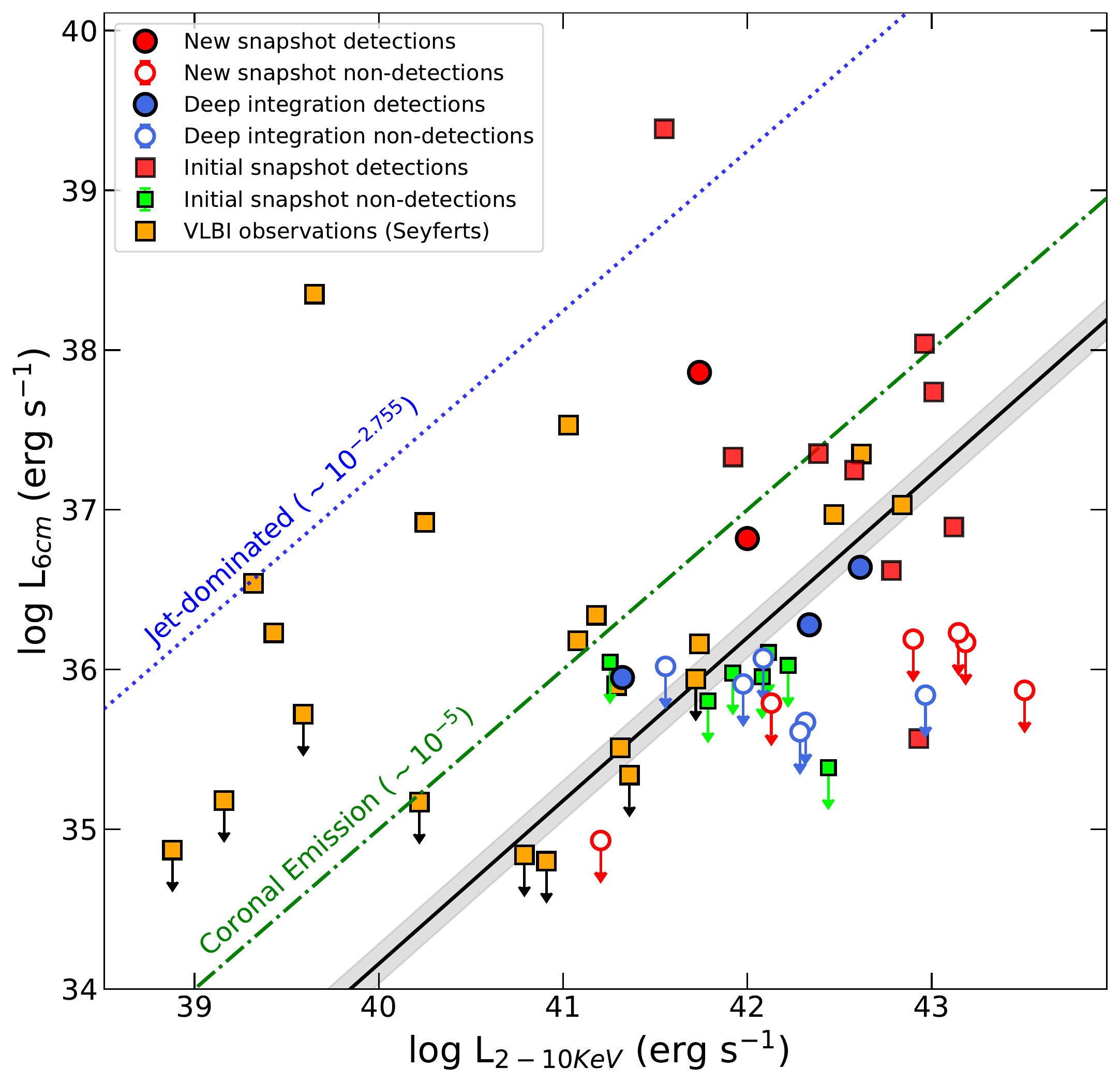}
\caption{6 cm radio peak luminosities compared with 2-10 keV X-ray luminosities for our total volume complete sample, and VLBI observations of a sample of local Seyfert galaxies from \citet{Panessa_Giroletti_2013MNRAS.432.1138P}. The blue dotted line indicates the separation between radio loud (RL) and radio quiet (RQ) AGNs as introduced by \citet{panessa_2007A&A...467..519P} and separates Seyferts from LLRGs of $L_\mathrm{R}$/$L_\mathrm{X}$ = 10$^{-2.755}$.
The green dash-dotted line indicates the well established relation for coronally active cool stars of~$L_\mathrm{R}$/$L_\mathrm{X}$ = 10$^{-5}$~\citep[][]{Guedel_1993ApJ...405L..63G}{}. The black solid line represents the fit line including our non-detection upper limits and the gray shading indicates the 1-$\sigma$ confidence region around the best fit line. }
\label{fig:radio_vs_soft_xray_comparison}
\end{figure}

In Figure 7 (left panel) of Paper I, we showed the relationship between X-ray (2-10 keV) and radio continuum (6 cm) luminosities for the initial volume-limited sample selected for FRAMEx together with a sample of local Seyfert galaxies from ~\citet[][]{Panessa_Giroletti_2013MNRAS.432.1138P}{} and Galactic and extragalactic black holes from ~\citet[][]{Merloni_2003MNRAS.345.1057M}{}. The main conclusion drawn from this figure was that the majority of the VLBA observed AGNs at parsec scales that were underluminous in radio as compared to the lower-resolution studies, found well below the ~$L_\mathrm{R}$/$L_\mathrm{X}$ = 10$^{-5}$
scaling relationship for pure coronal emission~\citep[][]{Guedel_1993ApJ...405L..63G,Laor_Behar_2008MNRAS.390..847L,panessa_2019NatAs...3..387P}{}. In this work, we recreated a similar figure (Figure~\ref{fig:radio_vs_soft_xray_comparison}) to explore this relationship further focusing only on the extragalactic sources. The higher sensitivity of the deeper 4-hour on-source integrations allowed for the recovery of three additional sources (blue filled circles), which were not previously detected in the 1-hour snapshot program (Paper I). The radio luminosities of these deeper observations are found to be lower compared to those obtained in previous snapshot detections from Paper~I (red squares), and in additional snapshot detections from this paper (red filled circles). In Figure~\ref{fig:radio_vs_soft_xray_comparison}, we plot the coronal emission relationship of~$L_\mathrm{R}$/$L_\mathrm{X}$ = 10$^{-5}$~\citep[][]{Guedel_1993ApJ...405L..63G}{} (green dash-dotted line) and the redefined radio-loudness parameter of ~$\log_{10} R_\mathrm{X} = \log_{10}(L_\mathrm{6~cm} / L_\mathrm{2-10~keV}) = -2.755$ from~\citet{panessa_2007A&A...467..519P} (blue dotted line), which is a more rigid threshold as compared to the classical divison of~$\log_{10} R_\mathrm{X} = -4.5$ between RL and RQ AGNs by~\citet{Terashima_2003ApJ...583..145T}. The latter division was derived based on a fixed ratio
between the monochromatic luminosities at radio and optical
frequencies (R $\equiv$ $L_\mathrm{6~cm} / L_\mathrm{B}$ = 10), and ~\citet[][]{panessa_2007A&A...467..519P}{} claimed that one should avoid the use of optical bands where absorption occurs naturally and might overestimate the value of R. The redefined boundary of the radio-loudness parameter was derived from comparing a sample of local Seyfert galaxies with a sample of low luminosity radio galaxies (LLRGs). Both the samples showed a similar correlation slope of $L_\mathrm{X} \propto L_\mathrm{R}^{0.97}$, but Seyfert galaxies were three orders of magnitude less luminous in the radio band than LLRGs. Both the samples exhibited two different distributions in R$_\mathrm{X}$, where the maximum
separation between them was located at~$\log_{10} R_\mathrm{X} = -2.755$. The majority of the LLRGs were found on the radio-loud side and a common non-thermal origin for radio and X-ray emission has been suggested (e.g., synchrotron radiation from a relativistic jet). On the other hand, the origin of radio-quieter Seyfert sample was attributed to the disk-corona system.

With our improved  measurements and additional sources, we find that radio luminosities range widely over 3 to 4 orders of magnitude. The black solid line represents the best fit to the data including the non-detections, which are treated as upper limits, obtained with the ASURV regression analysis package~\citep[]{Lavalley_1992BAAS...24..839L} and determined in the same way as Paper I. Our best-fit line yields log $L_\mathrm{6 cm}$ = (1.02) log $L_\mathrm{2-10 keV}$ $-$ 6.67 or $L_\mathrm{X} \propto L_\mathrm{R}^{0.98}$, which is similar to the correlation slope for radio quiet Seyferts found by~\citet[][]{panessa_2007A&A...467..519P}{}. However, our sample contains radio nuclei that are $\sim$2 orders of magnitude lower in radio luminosity.
Indeed, the higher-resolution VLBA data lie well below the radio-loud jet domination line, and only a few sources are close to the coronal line with the majority of our sample lying below. Our findings are broadly consistent with the results from Paper I and suggest that no significant coronal synchrotron radio emission is produced in these AGNs, or that the emission is attenuated such that it is not detectable at our sensitivity limits.

\subsection{The Dependence of the Fundamental Plane of Black Hole Activity and Radio-Loudness on the Accretion Rate}
\begin{figure*}
\centering
\includegraphics[trim=0mm 0cm 0mm 0cm, clip, scale=0.35]{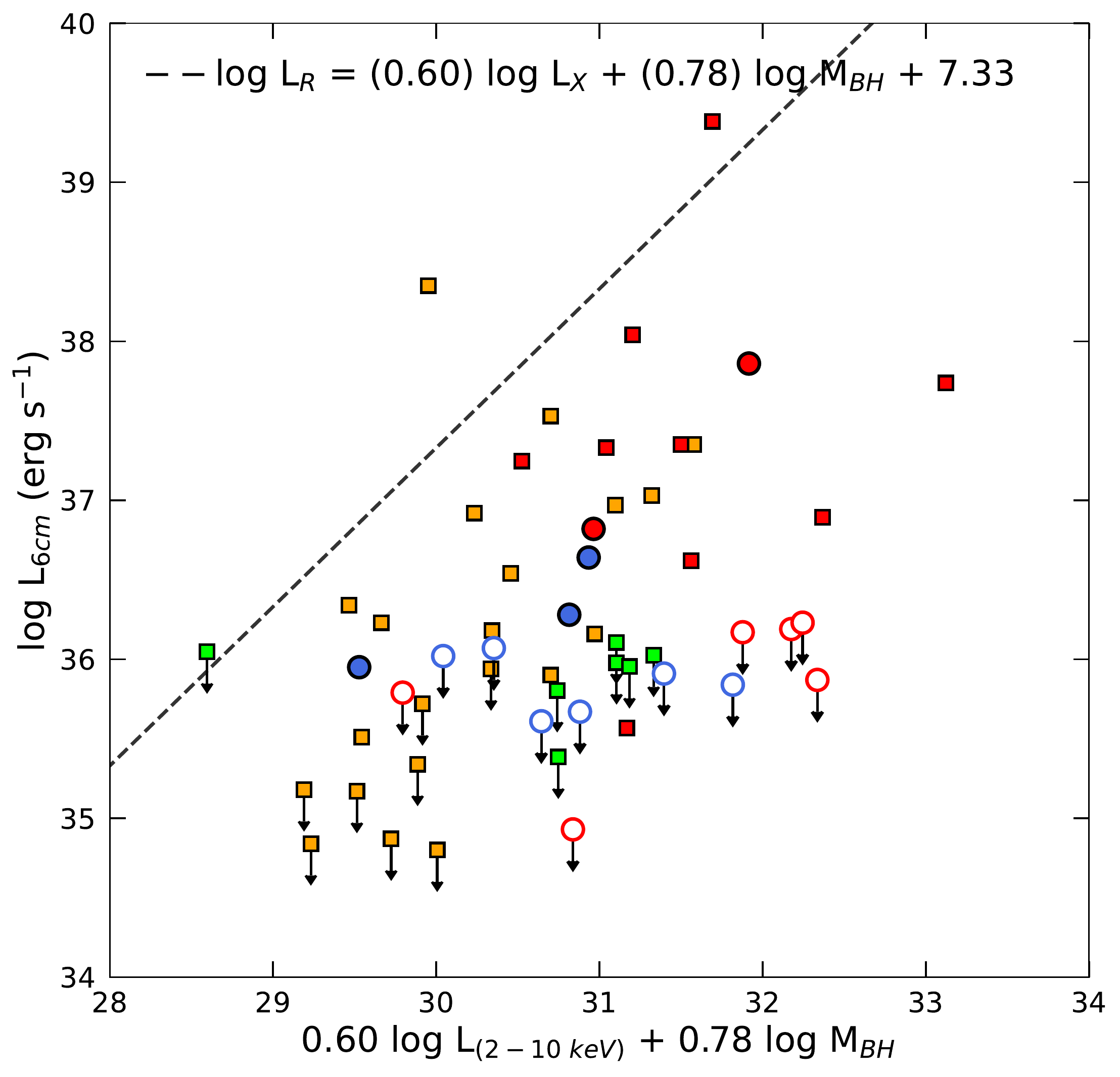}
\includegraphics[trim=0mm 0cm 0mm 0cm, clip, scale=0.35]{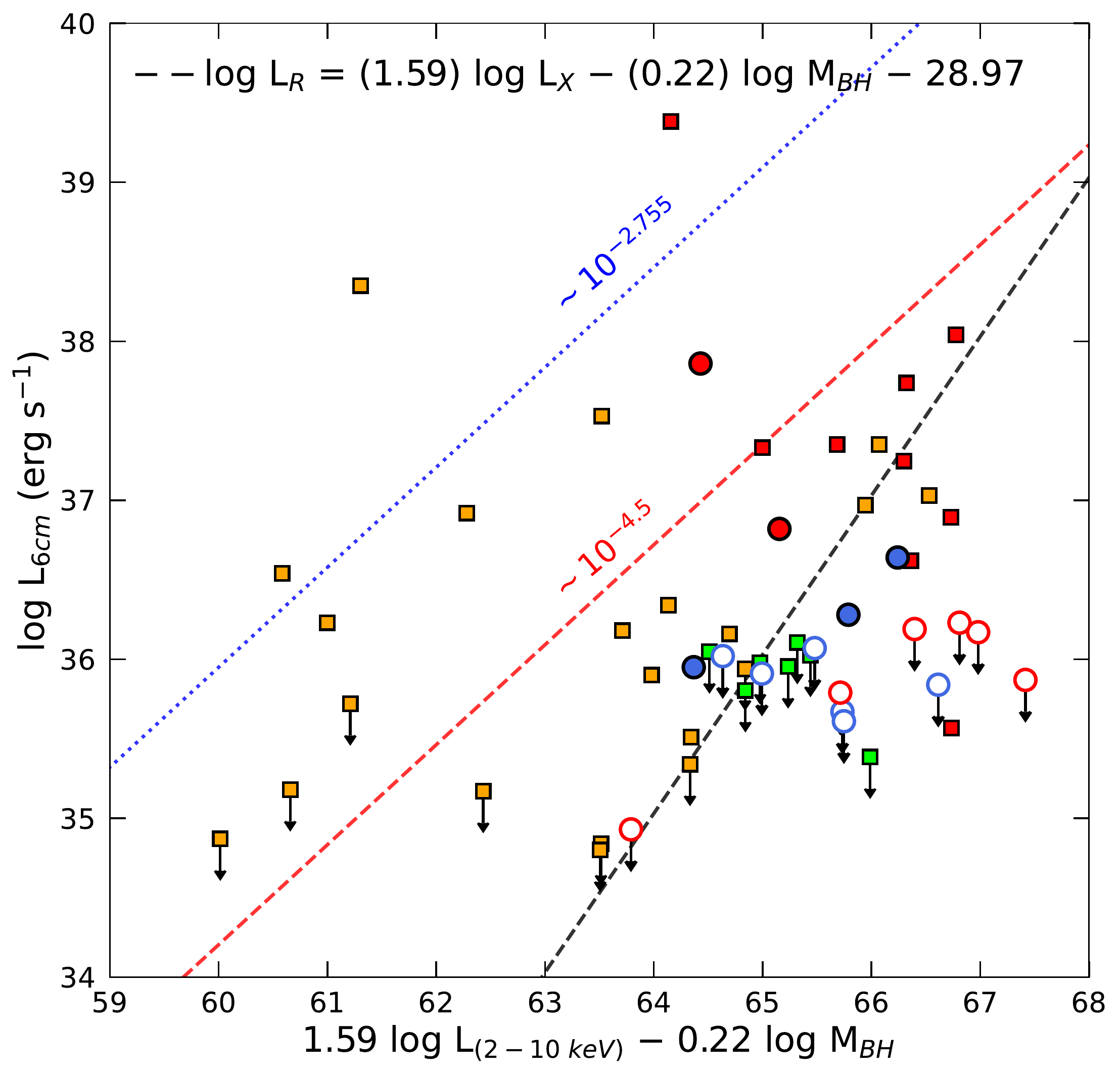}
\caption{The fundamental plane of black hole activity described by ({\it left:})~\citet{Merloni_2003MNRAS.345.1057M} and ({\it right:})~\citet{Dong_2014} using peak radio luminosities. The symbols are the same as those in Figure~\ref{fig:radio_vs_soft_xray_comparison}. The red dashed line on the right panel represents the classical division between RL and RQ AGNs by~\citet{Terashima_2003ApJ...583..145T}.}
\label{fig:FP_compare}
\end{figure*}

There is some evidence that supermassive black holes in AGNs behave similarly to stellar mass black holes in X-ray binaries (XRBs)~\citep[see, e.g.,][]{Done_2005MNRAS.364..208D,kording_2006MNRAS.372.1366K,McHardy_2006Natur.444..730M}. More specifically, different classes of AGNs show similarities in their X-ray and radio properties with XRBs in  different spectral states, which are
driven by the accretion rate \citep[see][for a review]{McClintock_2006csxs.book..157M}.  For this reason, it is important to compare the different AGN classes with XRBs in the appropriate spectral state. 
Since our volume complete sample is mostly made of sources optically classified as Seyfert, which are accreting at a moderately high rate,  it is probably more appropriate to compare it to the FP introduced by \citet{Dong_2014}  that is restricted to XRBs in their high-accreting state and bright AGNs. On the other hand, the \citet[][]{Merloni_2003MNRAS.345.1057M} FP used a more heterogeneous sample possibly dominated by low-accreting sources. However, both of these FP relations were derived from samples observed at relatively large physical scales, based on VLA and ATCA radio observations, which probe scales of hundreds to thousands of parsecs around the AGNs while VLBA/VLBI measurements probe much more localized sub-parsec scales.

Several other FP studies on parsec to tens of parsecs scales provide some useful context.  For example, \citet[][]{Saikai_2018A&A...616A.152S}{} observed a sample of 76 low-luminosity active galactic nuclei (LLAGN) with the VLA at 15 GHz, consisting of low-luminosity Seyfert galaxies, low ionization nuclear emission region (LINER) objects, and transition nuclei. This low-luminosity AGN (LLAGN) sample appears to follow a FP, however, differently from our study, the radio fluxes were extracted from larger regions compared to our milli-arcsecond study and the radio frequency used was 15 GHz (as opposed to our 5 GHz). In the recent work from \citet[][]{Gultekin_2019ApJ...871...80G}{}, the FP was refined using a sample restricted to objects with BH masses determined directly from dynamical methods. Differently from our work, \citet[][]{Gultekin_2019ApJ...871...80G}{} utilize lower radio resolution observations obtained with the VLA and different radio frequencies transformed into 5 GHz by assuming canonical single power-law spectra. However, very recently, \citet[][]{panessa_2022MNRAS.tmp.1693P}{} in a study based on radio frequencies ranging from 5 to 15 GHz, demonstrated that the radio spectra of radio quiet AGNs are described by different models including convex models and broken power laws, not only by single power laws (see Table 3 in~\citet[][]{panessa_2022MNRAS.tmp.1693P}{}). These findings suggest that the radio fluxes derived by \citet[][]{Gultekin_2019ApJ...871...80G}{} using a power-law canonical model should be taken with caution. 

Similarly to Figure 4 (left panel) of Paper I, we show our sample and the local Seyfert galaxies observed with VLBI from \citet[][]{Panessa_Giroletti_2013MNRAS.432.1138P} on the FP described by  \citet{Merloni_2003MNRAS.345.1057M} in the left panel in Figure \ref{fig:FP_compare} and in the right panel of the same figure we plot the same targets with the \citet[][]{Dong_2014} FP.  
Admittedly, a large scatter is present in both plots, but, with the exception of a few objects that are likely radio loud (above the classical division between RL and RQ AGNs \citep{Terashima_2003ApJ...583..145T}; red dashed line), our data appeared to be comparatively closer to the fundamental plane of black holes activity described by \citet[][]{Dong_2014}. In agreement with Paper~I, we conclude that any fundamental plane of black hole activity breaks down when the radio emission is resolved to parsec scales with no discernible correlation between black hole mass or X-ray luminosity, similar to the $L_R$/$L_X$ relation. 

\begin{figure}
\centering
\includegraphics[width =\columnwidth]{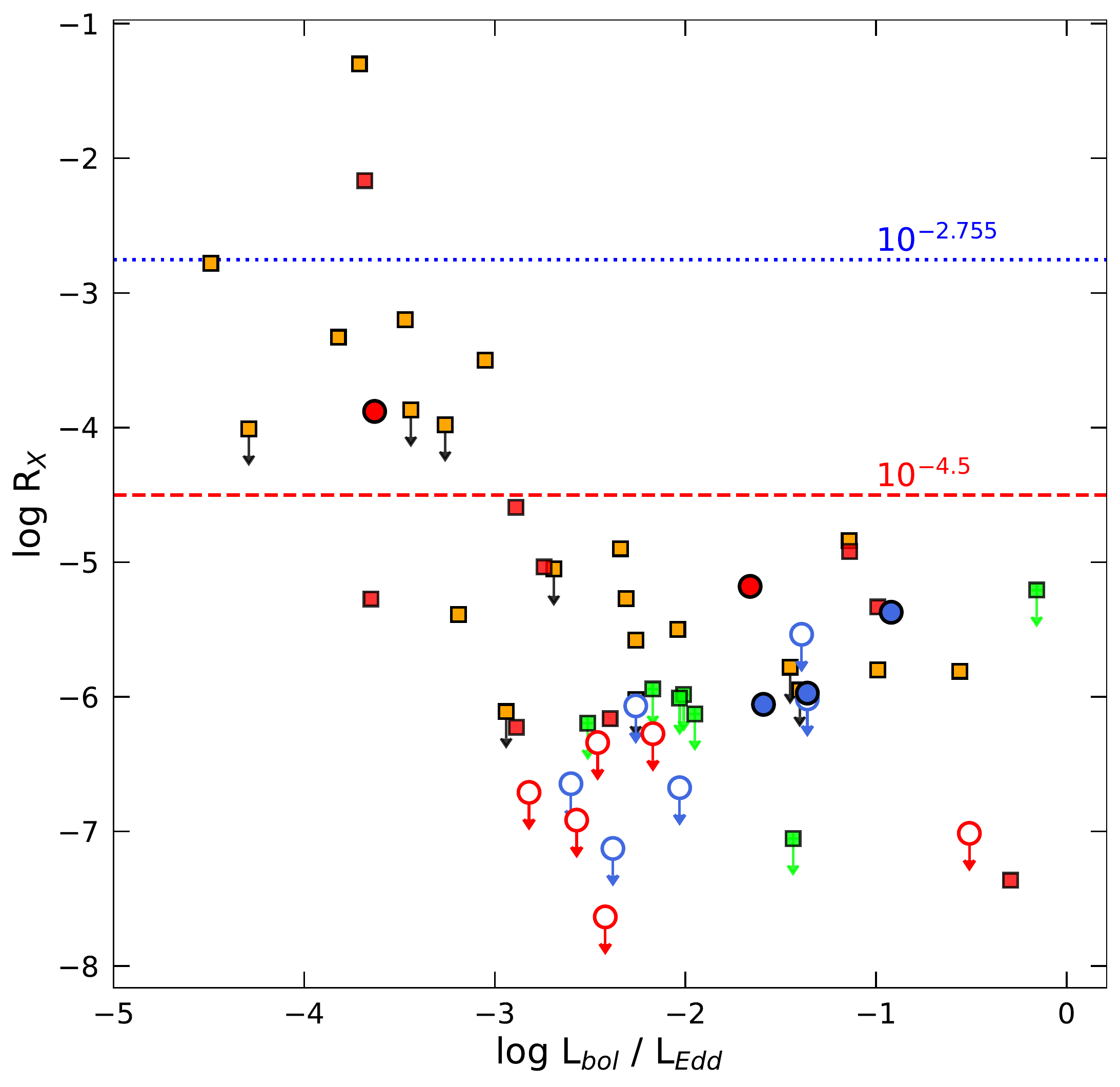}
\caption{Radio loudness vs Eddington ratio plot for our VLBA sample and the local Seyferts studied by~\citet[][]{Panessa_Giroletti_2013MNRAS.432.1138P}.  The symbols are the same as those in previous two figures.}
\label{fig:Rx}
\end{figure}

Interestingly, the radio loudness parameter,~$R_\mathrm{X} = L_\mathrm{6~cm} / L_\mathrm{2-10 keV}$, shows a negative trend when plotted as a function of the Eddington ratio $L_{bol}/L_{Edd}$, 
which parametrizes the source's accretion rate as shown in Figure~\ref{fig:Rx}. 
The figure shows $R_\mathrm{X}$ versus the Eddington ratio, with the \citet{panessa_2007A&A...467..519P} demarcation line $R_{X}$ = $-$2.755, shown along with the $R_{X}$ = $-$4.5 line representing the traditional separation between RL and RQ AGNs (which corresponds to a value of 10 for the optically-based radio-loudness parameter R). This is the trend observed in XRBs in their transition from the ``low/hard'' state (a low-luminosity state
with a hard, non-thermal X-ray spectrum), which is consistently associated with steady radio emission attributed to a jet, to the ``high/soft" state (a high luminosity
state with a soft, thermal X-ray spectrum) where the radio emission is suppressed. This inverse correlation was previously seen by \citet[][and the references therein]{Sikora_2007ApJ...658..815S}, using the traditional optically-based radio loudness parameter instead of the X-ray based one used here. This anti-correlation was also described by \citet[][]{Wang_2004ApJ...615L...9W}, expressing the idea of an inter-connection between radio jet and the Eddington ratio of the accretion disks. Both of these properties were found to be inversely correlated in a sample of 35 Blazars with VLBI observations. From Figure \ref{fig:Rx}, we find that most of our non-detections are grouped together in the highest accretion and lowest radio loudness region, the bottom right corner of the plot. The high accretion regime in black hole systems appears to disfavor the formation of radio emitting jets. This conclusion is also in agreement with the wind-jet inverse relation in radio-loud AGNs reported by \citet[][]{Mehdipur_2019A&A...625A..25M}, and may be a possible explanation for our radio quiet AGN sample. %

\subsection{Flux Variability and Absorption}

\begin{figure*}
\centering
\includegraphics[width=\columnwidth]{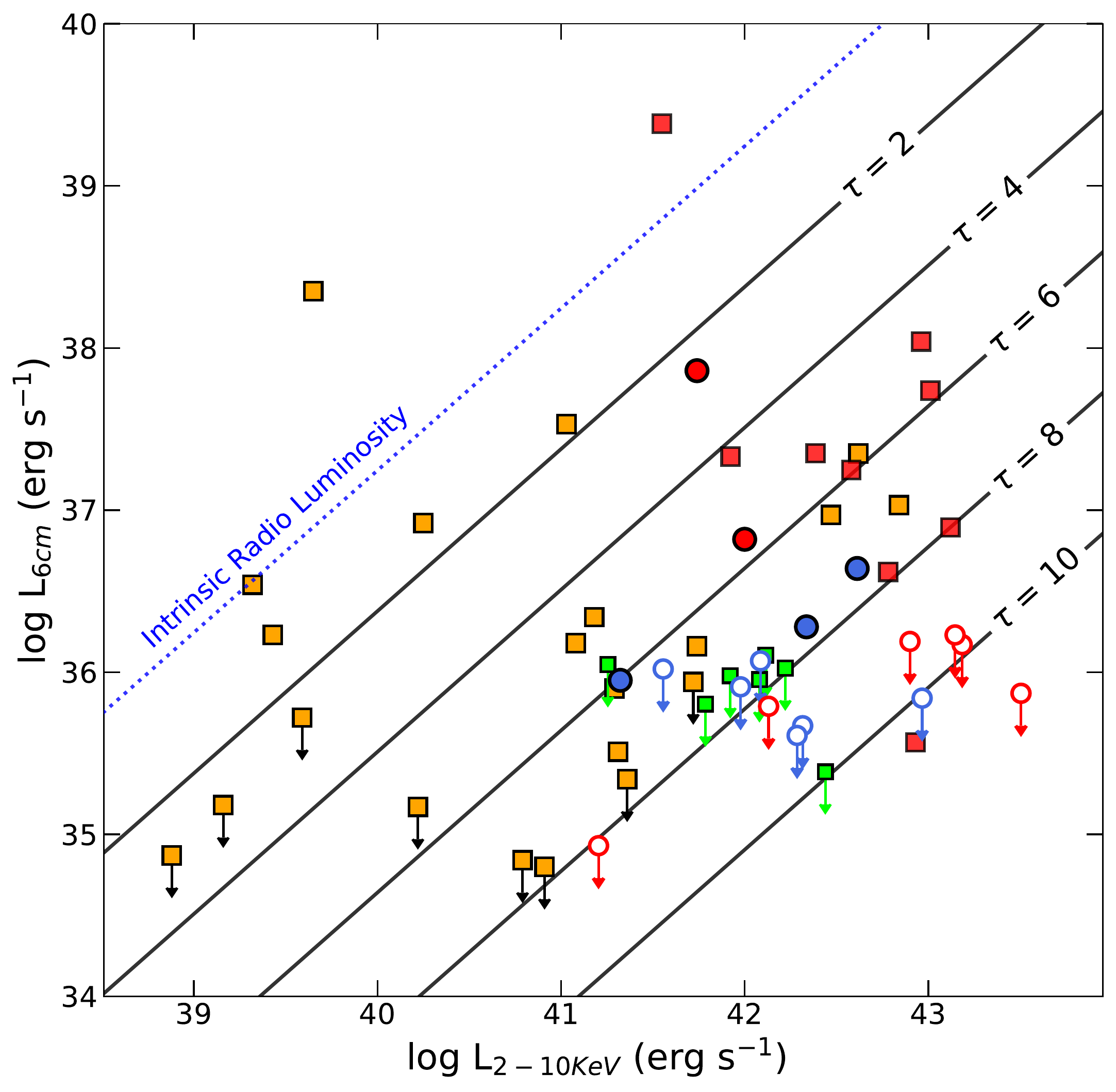} 
\includegraphics[trim=0mm 0cm 0mm 0cm, clip, scale=0.385]{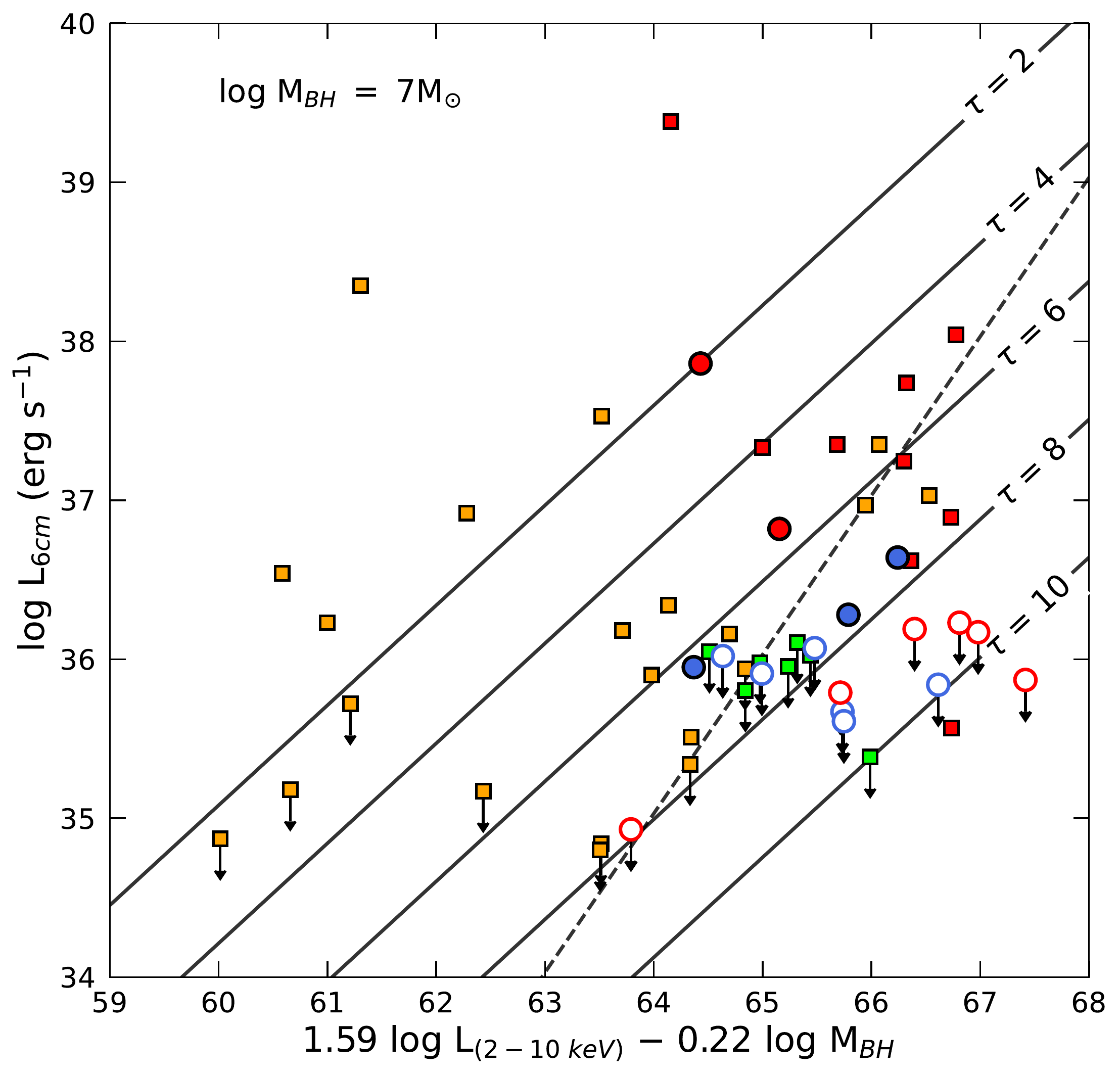}
\caption{{\it Left :} Similar comparison between 6 cm radio peak luminosities with 2-10 keV X-ray luminosities as shown in Figure~\ref{fig:radio_vs_soft_xray_comparison} and {\it Right :}  FP described by ~\citet{Dong_2014} as shown in the right panel of Figure~\ref{fig:FP_compare}, but different optical depth or opacity (tau, $\tau$) lines are drawn assuming that all AGNs have the same intrinsic radio/X-ray ratio and synchrotron self-absorption leads to the large dispersion seen.}
\label{fig:radio_vs_xray_comparison}
\end{figure*}

\begin{figure*}
\centering
\includegraphics[width=\columnwidth]{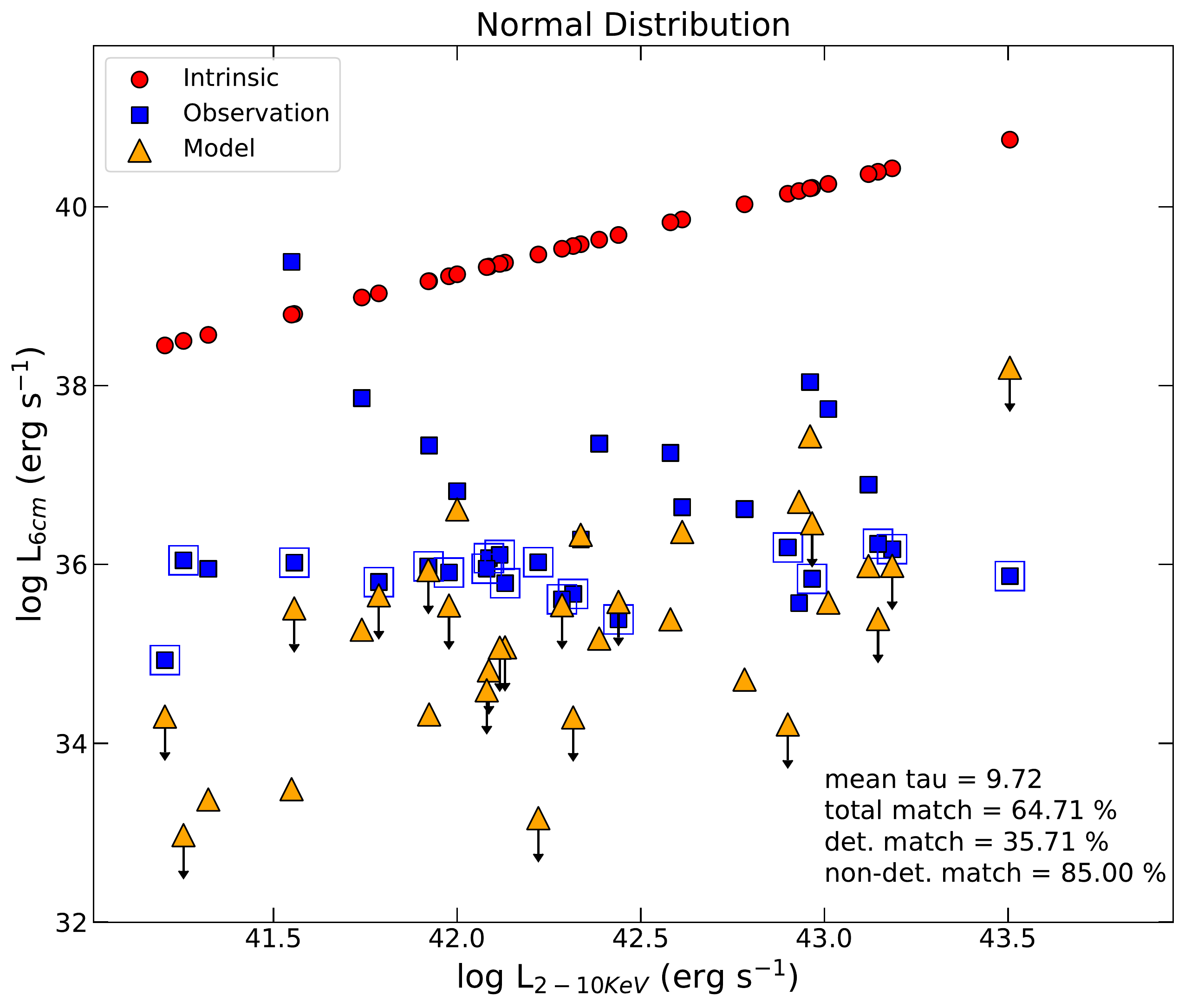} 
\includegraphics[width=\columnwidth]{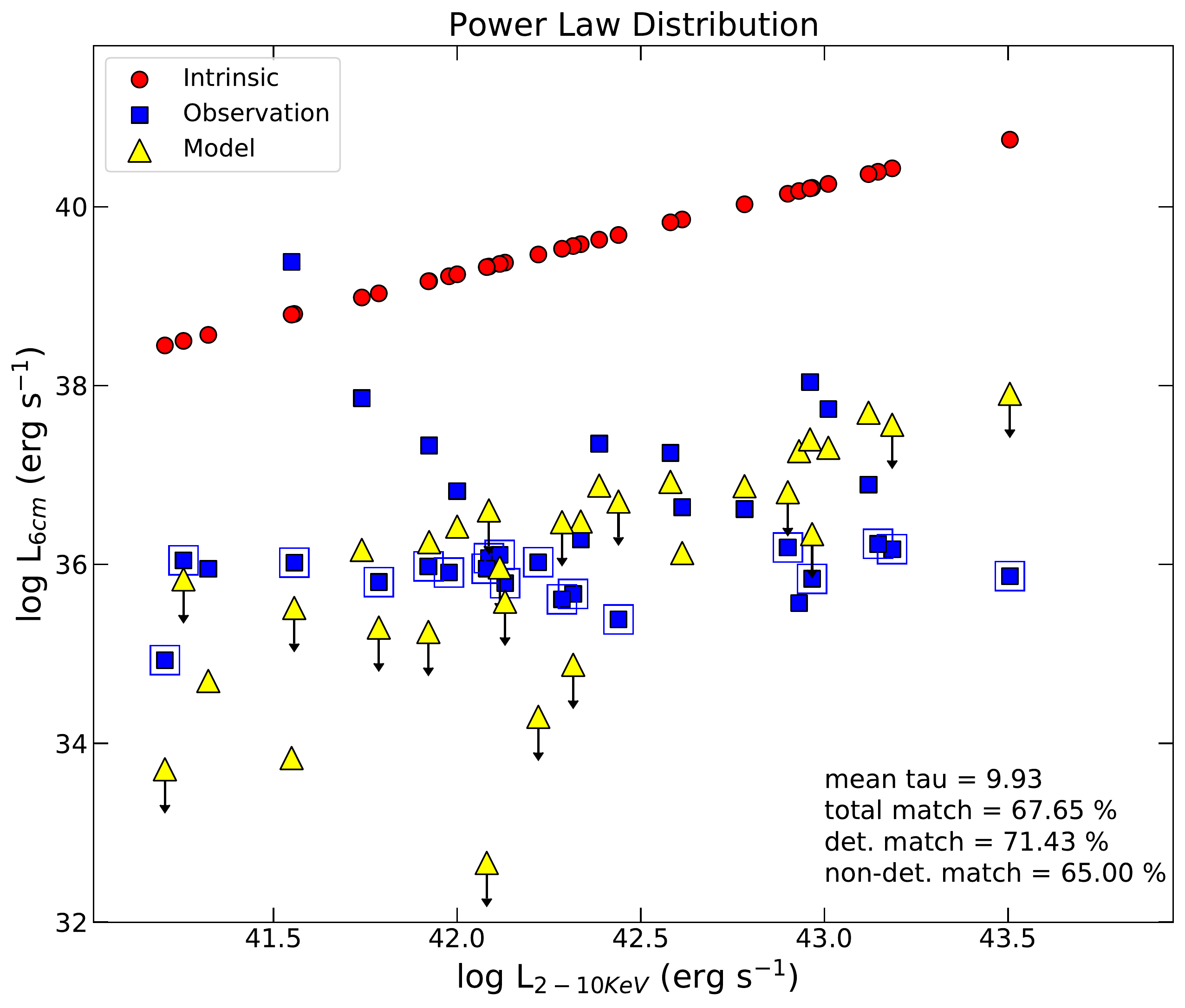} 
\caption{Using different distributions ({\it left :} Normal, {\it right :} Power law) of optical depth ($\tau$) values, we generated model observation sets to compare with our real VLBA observations. Intrinsic radio luminosities (red circles) were calculated from the jet dominated radio-to-X-ray relation (blue dotted line from Figure~\ref{fig:radio_vs_soft_xray_comparison}). The real observation sets are presented as blue squares where the points within boxes are non detections. The calculated model observation set are presented here in orange (Normal) and yellow (Power Law) triangles. The mean values of optical depth and the percentages of matching in total set, detection and non detection only are shown as well. With a similar mean value of $\sim$10, a random power law distribution of absorption (optical depth) parameter yields a better agreement with the observed detections ($\approx$ 71\%) and non-detections ($\approx$ 65\%) compared to a normal distribution.}
\label{fig:model_obs}
\end{figure*}


In FRAMEx Paper II~\citep[][]{Fernandez_2022ApJ...927...18F}, a nearby radio quiet AGN, NGC 2992, was observed simultaneously in X-ray (2-10 keV) and radio (5 cm) over a period of six months with an angular resolution similar to ours and showed a large drop in radio luminosity (over
a factor of $>$ 3) within a few months timescale. Interestingly, this radio dimming event was observed shortly ($\sim$ a month) after an increase in the X-ray 2$-$10 keV flux, indicating an increase of ejected materials near the source and higher opacity at radio frequencies by free-free absorption. Interestingly, the importance of self-absorbed synchrotron emission at~$\sim$GHz frequencies is predicted by the model proposed by~\citet[]{Ishibashi_2011A&A...525A.118I}, where shocks accelerate relativistic electrons within the accretion flow of radio quiet AGNs, the opacity decreases with the increase of the frequency and the transition from optically thick to optically thin regimes lies around a few tens of GHz. For our sources, which are observed at $\sim$5 GHz, we thus expect free-free or self-absorbed synchrotron radiation or a combination of both as one of the major reasons for such low radio luminosity values. Alternatively, we simply might have observed our radio quiet sources during a similar ``off'' state (drop in radio luminosity) seen in NGC 2992 during the 6 months monitoring campaign. 

To quantify the opacity of the absorber near our AGNs, we calculated new absorbed radio-X-ray correlation lines depending on the strengths of absorption or optical depth ($\tau$) values. We assumed that without any absorption at all, the AGNs would follow the jet dominated radio-X-ray correlation of $L_\mathrm{R}$/$L_\mathrm{X}$ = 10$^{-2.755}$, similarly to radio loud AGNs. In the left panel of Figure~\ref{fig:radio_vs_xray_comparison}, our calculated $\tau$ lines are shown on top of the $L_\mathrm{R}$/$L_\mathrm{X}$ plot for a range of optical depth values ($2 < \tau < 10$), where the higher the absorption, the larger the deviation is seen from the intrinsic radio luminosity line. Detected sources were found to be distributed over a broader optical depth range ($2 < \tau < 8$), whereas the non-detections are clustered around a higher optical depth range ($6 < \tau < 10$). NGC 1068 is the only detected source found beyond the $\tau$ = 8 line. In the right panel, we include optical depth lines to the FP plot described by ~\citet[][]{Dong_2014}{}, but for an average black hole mass of $10^{7}$~M$_{\odot}$. This is roughly the average of all black hole masses for our radio quiet AGN sample. The scatter seen in the plot with the the optical depth lines, demonstrates a similar pattern to the radio-to-X-ray correlation plot. 
In a simulation of sub-Eddington accretion disks around a supermassive black hole of mass $5\times10^8 ~{\mathrm{M_\odot}}$, \citet[][]{Jiang_2019ApJ...885..144J}{} showed that for an increase in the accretion rate, the disk becomes thicker and the total optical depth (the sum of absorption and scattering opacity) increases. 

Finally, using different random distributions of optical depths, we generated synthetic data points to be compared with the observed ones. In Figure~\ref{fig:model_obs}, we illustrate the comparison between our model set and real observation set for two different random distributions of optical depth. We found that a power law distribution with a mean optical depth $\sim$10, yields a better agreement with the observed detections ($\approx$ 71\%) and non-detections ($\approx$ 65\%) compared to a normal distribution. In particular, it appears that the power-law model shows a good agreement with the real data when  $8 < \tau < 12$. This result suggests that the presence of high optical depth regions producing absorption in radio emission may play a prominent role in reducing the rate of detections and in producing weak radiation. 
 

Throughout the paper, we have discussed different possible origins of radio emission from our sub-parsec scale observations, such as relativistic particles accelerated in shocks and winds or low-luminosity outflows. To shed some light on the dominant emission mechanism, we calculated the brightness temperature from the measured fluxes of our detected sources (see Table~\ref{tab:radio_xray_luminosity}). We find brightness temperatures of $10^{7.6}$~K and $10^{6.3}$~K from our two VLBA snapshot detections (NGC 3147 and NGC 3516), indicating either compact jet-associated or corona-associated synchrotron radiation as their likely dominant radio emission mechanism. This is in agreement with 
the correlation plot between the X-ray and sub-parsec scale radio emission in Figure~\ref{fig:radio_vs_soft_xray_comparison}. On the other hand, for the deep observation objects, the brightness temperature range of $10^{6}$~K$~ \geq T_b\geq10^{5.6}$~K suggests that thermal free-free emission from accretion disk winds/shocks may be the primary source of radio emission. Overall, our data are inconclusive in distinguishing between competing emission mechanisms.

\section{Conclusions} \label{sec:conclusions} 

In this work, we have expanded the investigation of the X-ray and radio properties of the original volume-limited  sample of AGNs presented in Paper I by adding VLBA deep integrations (4 hours per source) of 9 objects and including VLBA snapshot observations of 9 additional AGNs with declination limits of $-30\arcdeg$  and $+80\arcdeg$ and D~$\leq$~40 Mpc. Our three different X-ray data sets, which come from \emph{Swift}~BAT, \emph{Swift}~XRT and NuSTAR, have made it possible to derive good-quality broad-band X-ray spectra and absorption-corrected 2-10 keV luminosities that were compared with the respective radio luminosities and then used in the FP of BH activity for radiatively efficient radio-quiet AGNs. Our main findings are listed below:
{\begin{enumerate}
    \item With the help of improved, deeper VLBA radio observations ($\mathrm{RMS}\sim8~\mu$Jy), we recovered three sources that were not detected in our initial snapshot observations and we found two more detections from our additional snapshot observations with a similar observation depth  to our previous campaign ($\mathrm{RMS}\sim20~\mu$Jy). Despite the increased sensitivity of our observations, the majority of our sample is still undetected. Unlike strong jet-dominated radio loud AGNs, our sources lie well below the stringent threshold proposed by \citet[][]{panessa_2007A&A...467..519P}, as well as the $R_\mathrm{X}$ = $-$4.5 line commonly used in the literature. 
    \item We compared the radio luminosities as a function of X-ray emission from our radio-quiet AGN sample and found a correlation slope of $L_\mathrm{X} \propto L_\mathrm{R}^{0.98}$ for these high resolution radio observations. This is similar to the slope found by~\citet[][]{panessa_2007A&A...467..519P}{} for a sample of radio-quiet Seyfert galaxies but our sources were found to be $\sim$3 orders of magnitude less luminous in the radio. Consistent with our previous findings from Paper I, the~$L_\mathrm{R}/L_\mathrm{X}$ values are between $10^{-8}$ to $10^{-4}$ and the majority of our sample lies well below the fiducial $10^{-5}$ relationship for coronal synchrotron emission, showing no significant coronal synchrotron radio emission is produced in these AGNs. 
    
    \item In agreement with the results from our previous snapshot campaign, our work confirms that when radio fluxes are measured on sub-parsec scales, these radio-quiet AGN fall out of the FP, regardless of the relation chosen. 
    \item When the X-ray based radio loudness ~$R_{X}\equiv L_{R}/L_{X}$ is compared to the accretion rate~$\sim$~$L_{bol}/L_{Edd}$, an anti-correlation is revealed, in agreement with the findings of \citet[][]{Sikora_2007ApJ...658..815S} , as well as with those from \citet[][]{Panessa_Giroletti_2013MNRAS.432.1138P} based  on VLBI observations. Our sample sources are mostly clustered in the highest accretion and lowest radio loudness region. Similarly to the high accreting XRBs in their ``high/soft'' state where the radio emitting jets get suppressed, the radio-quietness of our sample sources can be interpreted as the absence of jets and presence of wind/shocks due to interactions between disk outflows and accretion flows.
    \item The ``turned off'' or ``silent'' state observed in a nearby radio-quiet AGN, NGC 2992, during a 6 month VLBA monitoring campaign by~\citet[][]{Fernandez_2022ApJ...927...18F}{} may explain the non-detections of our sources at such high angular resolutions. In addition to this, we expect free-free and self-absorption synchrotron radiation in $\sim$5 GHz frequency observations as one of the major reasons behind the low radio luminosities. We have shown that synthetic data obtained from a power-law distribution with optical depths of the order of 8$-$12 are consistent with our observed data plotted in the $L_{R}-L_{X}$ diagram shown in Figure~\ref{fig:radio_vs_soft_xray_comparison}. The high values inferred for $\tau$ support a scenario where the radio emission is produced in highly opaque outflows/winds, when the source accretes at a relatively high rate, whereas optically-thin radio jets are more likely produced by low-accreting systems.
\end{enumerate}}

Our deeper VLBA radio observations have allowed us to look into the sub-parsec regime of our sample of AGNs and has made it possible to study their core radio emission and some existing correlations which are still not well understood at these spatial scales. A future study of well-sampled radio spectral energy distributions can help us determine the true source of radio emission and disentangle contributions from potentially multiple components. Our next goal is to study high-resolution multi-wavelength observations at radio frequencies, as the radio spectral index, $\alpha$, will provide insights into a critical piece of the overall emission mechanisms of these radio quiet AGNs. 

\acknowledgments
This work supports USNO's ongoing research into the celestial reference frame and geodesy.\\

The National Radio Astronomy Observatory is a facility of the National Science Foundation operated under cooperative agreement by Associated Universities, Inc. The authors acknowledge use of the Very Long Baseline Array under the US Naval Observatory's time allocation. This work made use of data supplied by the UK Swift Science Data Centre at the University of Leicester.

\vspace{5mm}
\facilities{VLBA, Swift, VLA, EVLA}

\software{\textsc{aips}, Astropy~\citep{Astropy_2013A&A...558A..33A}, \textsc{casa}, \textsc{xspec}
          }


\appendix 
\section{Nustar Spectral Analysis} \label{appendix:a}
\setcounter{table}{0}
\renewcommand{\thetable}{\Alph{section}\arabic{table}}
\renewcommand*{\theHtable}{\thetable}

In the following, we summarize the specific details on the \nustar\ spectral analysis and \mbh\ determination of the individual sources. The main spectral parameters obtained by fitting this baseline model are reported in Table~\ref{tab:xray_spectra}. The X-ray scaling method was successfully applied to a sample of heavily obscured AGN, providing \mbh\ values consistent with those determined using disk maser measurements  \citep{gliozzi_2021MNRAS.502.3329G}.  The basic assumption of this method is that the Comptonization process producing the X-rays is the same in all black holes systems regardless of their mass and that the photon index $\Gamma$ is a faithful indicator of the accretion state of the source; specific details of this method are provided in \citet{gliozzi_2021MNRAS.502.3329G} and refences therein.\\

\noindent{\bf MCG-05-23-016} (obsid: 60001046006): The  \nustar\ spectrum of this bright Seyfert galaxy is well fitted by our baseline model, where the \texttt{MYTZ} component is substituted by a \texttt{zphabs} model.  The \mbh\ value derived from the X-ray scaling method $(1.4\pm0.5)\times10^7~{\mathrm{M_\odot}}$ is consistent within the respective uncertainties with the virial value of  $2.6\times10^7~{\mathrm{M_\odot}}$ obtained by \citet{Garcia_Bernete_2019MNRAS.486.4917G}.

\noindent{\bf NGC 2273} (obsid: 60001064002):  The \nustar\ spectrum of this Seyfert 2 galaxy is adequately fitted by our baseline model describing a Compton thick scenario, and  the \mbh\ value derived from the X-ray scaling method $(1.9\pm0.6)\times10^7~{\mathrm{M_\odot}}$ is broadly consistent with the  $(7.5\pm0.4)\times10^6~{\mathrm{M_\odot}}$ value obtained from mega-maser measurements by \citet{Kuo_2020MNRAS.498.1609K}, as explicitly demonstrated in the formal comparison between these two methods by \citet{gliozzi_2021MNRAS.502.3329G}.

\noindent{\bf NGC 3147} (obsid: 60101032002): This AGN has been classified as a true type 2 AGN, i.e., a source with an optical spectrum without any broad line and at the same time without any significant obscuration. The \nustar\ spectrum confirms that the X-ray primary emission is unabsorbed and hence the genuine lack of a BLR, which is thought to be a consequence of the very low accretion rate \citep{Bianchi_2019MNRAS.488L...1B}. Since the X-ray scaling method cannot be applied in this accretion regime, the small value derived  $(5.5\pm2.1)\times10^5~{\mathrm{M_\odot}}$ should not be considered reliable.

\noindent{\bf NGC 3516} (obsid: 60002042004): This source, optically classified as Seyfert 1.5 galaxy, is a changing-look AGN, which is characterized by extreme changes in  flux and spectrum. Since the X-ray spectral variability does not follow the standard softer-when-brighter trend typical of Seyfert galaxies, the \mbh\ value derived from the X-ray scaling method $(1.4\pm0.5)\times10^6~{\mathrm{M_\odot}}$ should be not considered reliable, and indeed it is about one order of magnitude smaller than the reverberation mapping value obtained by \citet{Feng_2021ApJ...909...18F}.

\noindent{\bf NGC 4102} (obsid: 60160472002): This source has been classified as a LINER but also as type 2 Seyfert galaxy.  The \nustar\ spectrum, which is well parametrized by a heavily absorbed but non Compton thick primary component, yields a fairly low \mbh\  of $(2.9\pm1.5)\times10^5~{\mathrm{M_\odot}}$.
Considering the possible LINER nature of the source and hence its intrinsically low accretion rate, in this case, we caution about the use of the  \mbh\  derived from the X-ray scaling method, because this method can only be applied to sources that are in the moderate or high accreting regime, which are characterized by the softer-when-brighter spectral transition, as shown by \citet{Jang_2014MNRAS.443...72J}.

\noindent{\bf NGC 5728} (obsid: 60662002002): The \nustar\ spectrum of this source confirms that it is Compton thick, as suggested by \citet{Comastri_2010ApJ...717..787C}. The  \mbh\ value derived from the X-ray scaling method $(8.3\pm2.9)\times10^6~{\mathrm{M_\odot}}$ is consistent with the constraints obtained by \citet{Kuo_2020MNRAS.498.1609K} based on mega-maser measurements.

\noindent{\bf NGC 7172} (obsid: 60061308002):  The \nustar\ spectrum of this Seyfert 2 galaxy is well parametrized by a Compton thin scenario. The  \mbh\ value derived from the X-ray scaling method $(1.03\pm0.35)\times10^7~{\mathrm{M_\odot}}$ is broadly consistent but lower than the value $5.5\times10^7~{\mathrm{M_\odot}}$ derived by \citet{Marinucci_2012ApJ...748..130M} utilizing the $M-\sigma_\star$ correlation. This seems to confirm the tendency of this correlation to overestimate the \mbh\ in type 2 AGN, as suggested by \citet{Ricci_2017MNRAS.471L..41R}.

\begin{deluxetable}{lcrrrrrHc} 
\setlength{\tabcolsep}{4pt}
\caption{NuSTAR X-ray Spectral Result }
\label{tab:xray_spectra}
\tablehead{\colhead{Source} & \colhead{Exposure} & \colhead{$\log(N_{{\textrm{H}},{\textrm{Bor}}})$}  & \colhead{$N_{{\textrm{H$_{MYTZ}$}}}$} & \colhead{$\Gamma$} & \colhead{$N_\textrm{BMC}$}  & \colhead{($\chi^2/$dof)} &  
&\colhead{{log($M_\mathrm{BH}$)}}\\
[-0.1cm]
\colhead{} &  \colhead{(ks)} & \colhead{~} &  \colhead{($10^{24}$ cm$^{-2}$)} & \colhead{~} & \colhead{~} & \colhead{~} &
& \colhead{($M_{{\textrm{$\odot$}}}$)}\\
[-0.1cm]
\colhead{(1)} & \colhead{(2)} & \colhead{(3)} & \colhead{(4)} & \colhead{(5)} & \colhead{(6)} & \colhead{(7)} & 
&  \colhead{(8)}}
\startdata
\hline
New Snapshots & & & & & & & \\
\hline
MCG-05-23-016 & $98$ & $23.91\pm0.01$ & $0.02\pm0.01$ & $1.95_{-0.01}^{+0.01}$ & $1.1_{-0.9}^{+0.7}\times10^{-3}$  & 1779.5/1707 & $95.4\pm0.2$ &  7.13\\
NGC 2273 & $23$ & $25.00\pm0.70$ & $6.80\pm0.40$ & $1.95_{-0.05}^{+0.05}$ & $3.1_{-0.2}^{+0.2}\times10^{-3}$  & 54/57& $36.0\pm13.0$ & 7.29\\
NGC 3147 & $49$ & \nodata~~~~ & $0.49\pm0.49$ & $1.75_{-0.03}^{+0.03}$ & $2.3_{-0.1}^{+0.2}\times10^{-5}$  & 217.2/270&   $2.8\pm0.4$  & \nodata~\\
NGC 3516 & $21$ & $23.84\pm0.04$ & $0.02\pm0.01$ & $1.88_{-0.01}^{+0.01}$ & $6.4_{-0.2}^{+0.2}\times10^{-5}$  & 726.9/647& $6.0\pm0.5$  & \nodata~\\
NGC 4102 & $21$ & $24.13\pm0.06$ &  $0.64\pm0.06$ & $1.53_{-0.07}^{+0.06}$ & $5.6_{-1.6}^{+2.7}\times10^{-5}$  & 167.1/176 & $9.2\pm0.3$ & \nodata~\\
NGC 5728 & $25$ & $24.30\pm0.04$ &  $1.09\pm0.04$ & $1.88_{-0.06}^{+0.05}$ & $4.5_{-0.9}^{+1.5}\times10^{-4}$ & 328.4/386&  $41.2\pm0.8$ & 6.91\\
NGC 7172 & $32$ & $24.14\pm0.03$ &  $0.10\pm0.01$ & $1.90_{-0.01}^{+0.02}$ & $6.7_{-0.6}^{+0.9}\times10^{-4}$ &  1155/1150& $83.8\pm0.8$& 7.01\\
UGC 6728 & $58$ & $24.06\pm0.06$ &  $0.010\pm0.002$ & $1.85_{-0.04}^{+0.03}$ & $1.6_{-0.4}^{+0.3}\times10^{-4}$ &  867.4/831& $14.4\pm0.2$ & 6.18\\
\hline
Deep Integrations & & & & & & & \\
\hline
NGC 1320 & $28$ & $24.50\pm0.09$ & $1.19\pm0.08$ & $1.70_{-0.07}^{+0.06}$ & $3.9_{-0.7}^{+0.7}\times10^{-5}$  & 73.6/61 & $5.4\pm0.5$ &  6.00\\
NGC 3081 & $55$ & $23.08\pm0.10$ & $0.79\pm0.01$ & $1.72_{-0.01}^{+0.01}$ & $4.9_{-0.2}^{+0.2}\times10^{-4}$  & 868.1/863& $65.7\pm6.6$ &  6.98\\
NGC 4388 & $21$ & $23.59\pm0.05$ & $0.45\pm0.04$ & $1.66_{-0.04}^{+0.04}$ & $3.3_{-0.4}^{+0.5}\times10^{-4}$  & 435.2/420&  $13.9\pm0.6$ &  6.63\\
NGC 4593 & $23$ & $25.50\pm0.60$ & $0.0002\pm0.0002$ & $1.88_{-0.01}^{+0.01}$ & $2.3_{-0.1}^{+0.1}\times10^{-4}$  & 671/679& $22.8\pm0.2$ &  6.59\\
NGC 6814 & $148$ & $24.26\pm0.04$ &  $0.008\pm0.001$ & $1.88_{-0.01}^{+0.01}$ & $3.0_{-0.1}^{+0.1}\times10^{-4}$  & 1598/1520 & $34.8\pm0.3$ & 6.25 \\
NGC 7314 & $100$ & $24.23\pm0.04$ &  $0.012\pm0.001$ & $2.03_{-0.01}^{+0.01}$ & $4.9_{-0.1}^{+0.1}\times10^{-4}$ & 1310.8/1276& $38.0\pm0.4$ &  6.11\\
NGC 7465 & $21$ & $23.83\pm0.06$ &  $0.01\pm0.01$ & $1.87_{-0.02}^{+0.02}$ & $1.2_{-0.1}^{+0.1}\times10^{-4}$ &  445.9/466& $12.6\pm0.2$  & 6.08\\
\enddata
\tablecomments{Columns: 1 = AGN name. 2 = \nustar\ FPMA exposure. 3 = column density calculated with the \borus\ model that parametrizes the continuum scattering and fluorescent emission line components. 4  = column density acting on the transmitted primary emission calculated with the \mytorus\ model;  for type 1 AGN the \texttt{phabs} is used instead. 5 = photon index. 6 = normalization of the BMC model.  7 = $\chi^2$ divided by degrees of freedom. 8 =  black hole mass.}
\end{deluxetable}

\noindent{\bf UGC6728} (obsid:  60376007002): In the best fit model of this Seyfert 1.2 galaxy the \texttt{MYTZ} component, which is more appropriate for heavily obscured AGN, was substituted by a \texttt{zphabs} model.  The  \mbh\ value derived from the \nustar\ spectral data, $(1.5\pm0.5)\times10^6~{\mathrm{M_\odot}}$ is consistent within the respective uncertainties with the reverberation mapping value $(7\pm4)\times10^5~{\mathrm{M_\odot}}$ \citep{Bentz_2016ApJ...831....2B}.

\noindent{\bf NGC 1320} (obsid: 60061036004): The best fit of this heavily absorbed Seyfert 2 galaxy was obtained by adding to our baseline model a Gaussian line at 6.4 keV (in the source's rest frame) and a scattered component produced by a putative optically thin ionized medium with a fraction of $\sim 4 \%$. The black hole mass derived with the X-ray scaling method  $(1.0\pm0.4)\times10^6~{\mathrm{M_\odot}}$ is broadly consistent  with the dynamical value ($5.3\times10^6 ~{\mathrm{M_\odot}}$) obtained from maser measurements, which however cannot considered as an accurate estimate because of the complex morphology and dynamics of the maser spots \citet{Gao_2017ApJ...834...52G}. For completeness, we also compared our \mbh\ estimate to the value ($1.3\times10^5 ~{\mathrm{M_\odot}}$) obtained using the most recent version of the Fundamental Plane for Black Hole Activity, based exclusively on objects with \mbh\ dynamically constrained  \citep{Gultekin_2019ApJ...871...80G}, which appears to underestimate \mbh\ by one order of magnitude.

\noindent{\bf NGC 3081} (obsid: 60561044002): The \nustar\ spectrum of this Seyfert 2 galaxy required two \borus\ components to properly fit the complex X-ray data. The  \mbh\ value derived from the X-ray scaling method $(9.5\pm3.6)\times10^6~{\mathrm{M_\odot}}$ is fully consistent with the value ($1.6\times10^7 ~{\mathrm{M_\odot}}$) obtained from gas dynamics \citep{Beifiori_2012MNRAS.419.2497B} and inconsistent with the Fundamental Plane for Black Hole Activity estimate ($2.8\times10^4 ~{\mathrm{M_\odot}}$), which underestimates \mbh\ by nearly three orders of magnitude.

\noindent{\bf NGC 4388} (obsid: 60061228002): The \nustar\ spectrum of this Seyfert 2 galaxy, one of the few with a disk maser measurement, was fitted with a scattered component (with a fraction of $\sim 17 \%$) added to our baseline model. The  \mbh\ value derived from the X-ray scaling method $(4.3\pm1.6)\times10^6~{\mathrm{M_\odot}}$ is in agreement with the dynamical value derived from maser measurements \citep{kuo_2011ApJ...727...20K},  as explicitly shown in \citet{gliozzi_2021MNRAS.502.3329G}. The estimate based on the Fundamental Plane for Black Hole Activity ($2.1\times10^5 ~{\mathrm{M_\odot}}$) one more time severely underestimates the \mbh.

\noindent{\bf NGC 4593} (obsid: 60001149002): In the best fit model of this Seyfert 1 galaxy the \texttt{MYTZ} component, which is more appropriate for heavily obscured AGN, was substituted by a \texttt{zphabs} model. The  \mbh\ value derived from the X-ray scaling method $(3.9\pm1.3)\times10^6~{\mathrm{M_\odot}}$ is broadly consistent with the reverberation mapping \mbh\ estimate ($(9.8\pm2.1)\times10^6 ~{\mathrm{M_\odot}}$) obtained by \citet{Denney_2006ApJ...653..152D}, as well as the model-independent value ($(5.8\pm2.1)\times10^6 ~{\mathrm{M_\odot}}$) derived from X-ray variability\citep{middei_2019MNRAS.483.4695M}, whereas it is inconsistent with with the Fundamental Plane for Black Hole Activity value ($3.5\times10^5 ~{\mathrm{M_\odot}}$).

\noindent{\bf NGC 6814} (obsid: 60201028002): Similarly to the other Seyfert 1 galaxy of this sample, the \texttt{zphabs} model was used to parametrize the intrinsic absorption acting on the transmitted primary X-ray emission. Once more the \mbh\ value derived from the X-ray scaling method $(1.8\pm0.6)\times10^6~{\mathrm{M_\odot}}$ is fully consistent with the reverberation mapping value ($2.34\times10^6 ~{\mathrm{M_\odot}}$) obtained by \citet{pancoast_2014MNRAS.445.3073P}. Again the value derived from the Fundamental Plane for Black Hole Activity ($4.4\times10^4 ~{\mathrm{M_\odot}}$) substantially underestimates \mbh.

\noindent{\bf NGC 7314}  (obsid:  60201031002):The \nustar\  spectrum of this source which has been optically classified as Seyfert 1.9 but also as a Narrow Line Seyfert 1 galaxy because of its pronounced variability, is well fitted by our baseline model with the addition of a gaussian line at 6.38 keV (in the source's rest frame) and a  \texttt{zphabs} model instead of \texttt{MYTZ} . Because  of the relatively steep photon index, the 2005 decaying outburst of GRO J1655-40 cannot be used to determine \mbh. Instead, we estimated the black hole mass using all the other available patterns described in  \citet{gliozzi_2011ApJ...735...16G} (the rising phase of the 2005 outburst of GRO J1655-40, the 2003 decaying phase and the 2004 rising phase of GX 339-4, and the rising phase of the 1998 outburst of XTE J1550-564) and then computed the average value. The \mbh\ value derived with the X-ray scaling method $(1.3\pm0.5)\times10^6~{\mathrm{M_\odot}}$ is consistent with the virial estimate ($1.74\times10^6 ~{\mathrm{M_\odot}}$) obtained by \citet{Onori_2017MNRAS.468L..97O} by measuring NIR lines in the BLR and assuming a constant virial factor of 4.31, whereas it is inconsistent with with the Fundamental Plane for Black Hole Activity value ($3.9\times10^4 ~{\mathrm{M_\odot}}$).
 
\noindent{\bf NGC 7465} (obsid: 60160815002):The \nustar\ spectrum of this Seyfert 2 galaxy is reasonably well fitted by our baseline model without any additional component. The \mbh\ value derived from the X-ray scaling method $(1.2\pm0.4)\times10^6~{\mathrm{M_\odot}}$ is in general agreement with the  the virial estimate ($3.47\times10^6 ~{\mathrm{M_\odot}}$) obtained by \citet{Onori_2017MNRAS.468L..97O}, whereas the value derived from the Fundamental Plane for Black Hole Activity value ($1.7\times10^5 ~{\mathrm{M_\odot}}$) again underestimates \mbh\ by more than one order of magnitude.


\bibliography{radio_quiet_agn}{}
\bibliographystyle{aasjournal}



\end{document}